\documentclass{article}
\usepackage{arxiv}
\usepackage{cite}
\usepackage[T1]{fontenc}
\usepackage{amssymb}
\usepackage{amsfonts}
\usepackage{amsmath}
\usepackage{tikz}
\usepackage{pgfplots}
\usepackage{subfigure}
\usepackage{mathtools}
\usepackage{array}
\usepackage{balance}
\usepackage{xspace}
\usepackage{url}
\usepackage{textcomp}
\usepackage{algorithm}
\usepackage{algorithmicx}
\usepackage{algpseudocode}
\usepackage{lmodern}
\usepackage[british]{babel}
\usetikzlibrary{positioning}
\pgfplotsset{compat=1.15}
\algrenewcommand\alglinenumber[1]{\scriptsize #1:}

\newcommand{\sys}{Saguaro\xspace}

\newcommand{\req}{{\sf \small request}\xspace}

\newcommand{\one}{{\sf \small propose}\xspace}

\newcommand{\THREE}{\textsf{COMMIT}\xspace}
\newcommand{\three}{{\sf \small commit}\xspace}
\newcommand{\PRE}{\textsf{PREPARE}\xspace}
\newcommand{\pre}{{\sf \small prepare}\xspace}
\newcommand{\PRED}{\textsf{PREPARED}\xspace}
\newcommand{\PREDQ}{\textsf{PREPARED-QUERY}\xspace}
\newcommand{\pred}{{\sf \small prepared}\xspace}

\newcommand{\abort}{{\sf \small abort}\xspace}
\newcommand{\ACK}{\textsf{ACK}\xspace}
\newcommand{\ack}{{\sf \small ack}\xspace}

\newcommand{\reply}{{\sf \small reply}\xspace}
\newcommand{\block}{{\sf \small block}\xspace}

\newcommand{\CMTQ}{\textsf{COMMIT-QUERY}\xspace}
\newcommand{\cmtq}{{\sf \small commit-query}\xspace}

\newcommand{\HIS}{\textsf{STATE}\xspace}
\newcommand{\his}{{\sf \small state}\xspace}
\newcommand{\HISQ}{\textsf{STATE-QUERY}\xspace}
\newcommand{\hisq}{{\sf \small state-query}\xspace}

\newtheorem{metalemma}{Lemma}[section]

\newtheorem{Lemma}[metalemma]{Lemma}
\newtheorem{Property}[metalemma]{Property}

\newenvironment{prop}{\begin{Property}\em}{\end{Property}}
\newenvironment{lmm}{\begin{Lemma}\em}{\end{Lemma}}

\newenvironment{prf}{\noindent{\bf Proof:}\rm}

\newif\ifextend
\extendtrue

\newif\ifnextend
\nextendfalse

\title{\sys: An Edge Computing-Enabled Hierarchical Permissioned Blockchain}

 \author{
 Mohammad Javad Amiri$^1$ \quad Ziliang Lai$^2$ \quad Liana Patel$^1$ \\
 {\bf Boon Thau Loo$^1$ \quad Eric Lo$^2$ \quad Wenchao Zhou$^3$}\\
$^1$Department of Computer and Information Science, University of Pennsylvania\\
$^2$Department of Computer Science and Engineering, Chinese University of Hong Kong\\
$^3$Department of Computer Science, Georgetown University\\
$^1$\{mjamiri, lianap, boonloo\}@seas.upenn.edu, $^2$\{zllai, ericlo\}@cse.cuhk.edu.hk, $^3$wzhou@cs.georgetown.edu
\vspace{2em}
}

\begin{document}

\maketitle

\begin{abstract}
We present {\em \sys}, a permissioned blockchain system designed specifically for edge computing  networks.
\sys leverages the hierarchical structure of edge computing  networks
to reduce the overhead of wide-area communication by presenting several techniques.
First, \sys proposes coordinator-based and optimistic protocols to process cross-domain transactions with low latency where the
lowest common ancestor of the involved domains coordinates the protocol or detects inconsistency.
Second, data are collected over hierarchy enabling higher-level domains to aggregate their sub-domain data.
Finally, transactions initiated by mobile edge devices are processed without relying on high-level fog and cloud servers.
Our experimental results across a wide range of workloads demonstrate
the scalability of \sys in supporting a range of cross-domain and mobile transactions.
\end{abstract}
\section{Introduction}\label{sec:intro}

Recent trends in edge computing present both new challenges and opportunities for distributed applications \cite{al2017technologies}\cite{li2020auditing}.
In the edge computing paradigm, computing shifts closer to the edge of the network \cite{hu2020heterogeneous}\cite{mach2017mobile}\cite{shi2016edge}.
Edge devices communicate in a peer-to-peer fashion in small geographic regions known as {\em spatial domains},
and can communicate with edge servers, fog servers,
and finally to cloud servers in a {\em hierarchical} fashion \cite{shafique2020internet}\cite{shi2016edge}\cite{tong2016hierarchical}.
The characteristics of edge networks have led to a wide range of distributed applications being proposed \cite{loghin2020disruptions},
e.g., intelligent transportation \cite{din20195g}, industry automation \cite{saad2019vision} and cross-border payments~\cite{cross-boader}.
Many of these applications require immutability, provenance, or verifiability over wide-area networks.
While point solutions exist, no general-purpose abstraction provides these capabilities in a unified manner.

Blockchain is a promising technology to realize the full potential of edge computing
by providing a common substrate usable by all edge computing-enabled applications \cite{dorri2016blockchain}\cite{dorri2017blockchain}\cite{xiong2018mobile}.
Increasingly, emerging uses of blockchains, in particular, {\em permissioned} blockchains,
require transaction processing over wide-area networks among a set of mutually distrustful known entities.
While distributed applications, e.g., contact tracing \cite{peng2021p2b}, crowdworking \cite{amiri2021separ},
supply chain management~\cite{amiri2019caper}\cite{amiri2022qanaat}\cite{tian2017supply}, and federated learning \cite{peng2021vfchain},
benefit from the unique features of permissioned blockchains, 
practical deployment of edge computing-enabled blockchain applications over wide-area networks
remains an elusive goal \cite{loghin2020disruptions}.

Traditional approaches for scaling distributed systems do not apply well over wide-area networks.
While sharding~\cite{corbett2013spanner} is used to partition data into multiple shards maintained by different clusters of machines,
blockchain sharding, independent of the deployment scenario, backfires over a wide area due to
the significant overhead of cross-shard transactions.
At one end of the spectrum, flattened permissioned blockchains~\cite{amiri2021sharper}
run a consensus protocol among all nodes of involved shards to process cross-shard transactions,
resulting in several messages crisscrossing high-latency low bandwidth links over the Internet.
On the other end of the spectrum, coordinator-based approaches~\cite{dang2018towards} do not fare much better, as the coordinator node is either close to clients or the data shards, which will not avoid slow network links when cross-shard transactions take place.
Trying to avoid wide-area transactions by replicating the entire ledger on every cluster,
e.g., GeoBFT~\cite{gupta2020resilientdb},
also merely shifts the wide-area communication from running the consensus protocol across data centers
to ledger synchronization messages over a wide-area network.
Moreover, current approaches do not address the mobility of nodes where
a mobile edge device temporarily migrates out of its local home domain to a remote domain and initiates transactions in the remote domain.

In this paper, we present {\em \sys},
a permissioned blockchain system
that leverages the hierarchical structure of edge computing infrastructures to support applications over a wide area.
At a high level, in \sys, nodes are organized in a hierarchical structure
from edge devices ({\em height$-$0}) to edge, fog, and cloud servers.
Nodes at each level are further clustered into fault-tolerant {\em domains} where domains might follow different failure models, i.e., crash and Byzantine. 
In \sys, each height$-$1 domain (i.e., edge servers) maintains its own blockchain ledger,
executes transactions received from their child edge devices (in parallel to other height$-$1 domains),
constructs its ledger and propagates the ledger to higher-level domains.
This hierarchical approach localizes network traffic for consensus and replication within local networks,
reducing wide-area communication overhead significantly.

\sys leverages the hierarchical structure of edge computing networks to achieve four main purposes.
First, \sys relies on the {\em lowest common ancestor} of all involved domains in the hierarchical structure
(i.e., a higher-level domain with minimum total distance from the involved domains) to 
process cross-domain transactions in a coordinator-based fashion with low latency.
Since edge servers execute transactions,
the load on the internal domains, e.g., cloud servers, is highly reduced, making \sys suitable for edge networks.

Second, this hierarchical structure enables height$-$2 and above domains to maintain
only a {\em summarized view} (e.g., selected columns or aggregated values) of their child domain ledgers.
In \sys, edge servers  order and execute transactions and periodically,
propagate the results to higher-level domains.
While height$-$1 domains maintain transactions in linear ledgers,
summarized ledgers at higher-level domains are structured as directed acyclic graphs
to capture dependencies resulting from cross-domain transactions.
These summarized views enable higher-level domains to
perform aggregation functions over their sub-domains data,
e.g., the total amount of exchanged assets in a micropayment application.

Third, the propagation of transactions through hierarchy enables \sys to optimistically
process cross-domain transactions.
Each involved height$-$1 domain of a cross-shard transaction
orders the transaction independently
without running costly cross-domain consensus protocols across height$-$1 domains, and then executes the transaction speculatively.
In case of any ordering inconsistencies, the higher-level domains and eventually the lowest common ancestor of the involved domains detect the inconsistency.

Finally, the hierarchical structure enables 
\sys to efficiently support the mobility of nodes without relying on high-level fog and cloud servers.
Mobile edge devices initiate transactions in different domains far from their initial local domain
while \sys establishes {\em mobile} consensus by sharing a node's state only between the local and remote domains.

\sys makes three key technical contributions:

\begin{itemize}
\item \sys supports data aggregation over hierarchy where 
transactions are executed and maintained in linear ledgers of height$-$1 domains while
above domains maintain 
only a {\em DAG-structured summarized view} of child domains.

\item  A suite of consensus protocols is provided to process transactions within and across
fault-tolerant domains.
\sys benefits from the hierarchical structure of edge networks for
the geographically optimized processing of cross-domain transactions using coordinator-based and optimistic protocols.

\item \sys supports mobility of nodes by providing a {\em mobile} consensus protocol
where edge devices initiate transactions in different domains far from their initial local area.
\end{itemize}

We validated these technical innovations by developing a prototype of \sys,
where our evaluation results across a wide range of workloads demonstrate
the effectiveness of \sys in scalably support a range of cross-domain and delay-tolerant transactions. 

% The rest of this paper is organized as follows.
% Section~\ref{sec:back} motivates \sys.
% The \sys model is introduced in Sections~\ref{sec:model}.
% Sections~\ref{sec:cons}, \ref{sec:propagation}, \ref{sec:opt} and \ref{sec:mobile} present transaction processing in \sys.
% Section~\ref{sec:exp} evaluates the performance of \sys.
% Section~\ref{sec:related} discusses related work, and
% Section~\ref{sec:conc} concludes the paper.

\section{Background and Motivation}\label{sec:back}

In an edge network, machines (i.e., devices, servers) are organized in a hierarchical structure where at the leaf level,
edge devices within a local area are connected to each other and
to an edge server domain (as the parent vertex).
Nearby edge servers (e.g., campus area)
are then connected to a fog server (e.g., metropolitan area) and finally, at the root level, cloud servers are placed \cite{tong2016hierarchical}.
The hierarchy might include multiple layers of edge, fog, or cloud servers.

We briefly describe several emerging applications that can realize the full potential of edge computing,
and yet require technology innovations by \sys to make this a reality.

\noindent
{\bf Accountable ridesharing and gig economy.}
In ridesharing applications, drivers give rides
to travelers through platforms, e.g., Uber and Lyft.
A ridesharing task usually occurs within a single domain (i.e., a local area).
However, supporting the mobility of cars across domains is challenging as
a driver registered in a local area might temporarily move
to another domain and give rides to travelers in that area.
Furthermore, a ridesharing application needs to aggregate specific data attributes from different spatial domains, e.g.,
the total number of tasks performed per day.
The aggregated data is needed for data analysis purposes and, more importantly, for satisfying global regulations, e.g.,
the total work hours of a driver, who might work for multiple platforms,
may not exceed $40$ hours per week to follow the
\emph{Fair Labor Standards Act}\cite{flsa}.
While the transparency and immutability of blockchains will aid in enforcing global regulations \cite{amiri2022prever}\cite{amiri2021separ},
permissioned blockchain solutions today are unable to work at a global scale given that the leading ridesharing firms are all globalized. 

\sys as a permissioned blockchain system can address the challenges above.
First, in \sys, each height$-$1 domain (i.e., edge servers) processes ridesharing tasks initiated by edge devices within a local area.
Second, within the hierarchical structure of \sys, while edge server domains process tasks and maintain the full record of transactions,
an aggregate version of records, e.g., the travel time or working hour attribute,
might be maintained by internal domains resulting in improved performance and enhanced privacy.
Third,  \sys efficiently addresses the mobility of edge devices across spatial domains.
Beyond ridesharing, the ability to add accountability and verifiable global statistics collection at Internet-scale
can be generally applied to any other mobile gig economy job.

\noindent
{\bf Micropayment.}
Most popular micropayment infrastructures do not allow users to do cross-application payments, e.g.,
an Apple Pay sender cannot send money to a PayPal receiver.
However, a hierarchical permissioned blockchain system can facilitate such micropayments.
For micropayments within the same spatial domain
and application domain (e.g., Alice pays Bob at a coffee shop, both using Apple Pay),
transactions can be committed efficiently and securely within the spatial domain.
For micropayments under the same spatial but different application domains
(e.g., Alice pays Bob in the same coffee shop, but Alice is using Apple Pay while Bob is using PayPal),
transactions can be executed efficiently if each edge server
hosts ledgers from different payment companies and executes the cross-domain transactions at the edge. 
For micropayments that cross spatial and application domains 
(e.g., Alice in the West pays Bob in the East), 
transactions can also be executed efficiently when ledgers are deployed in the entire wide-network hierarchy, but (cross-domain) consensus is established only among the involved domains. Finally, Alice and Bob may be on the move while micropayments are happening.
\sys aims to support such mobile micropayments as well.

\noindent
{\bf Resource management and provisioning.}
As mobile devices and Internet-of-Things (IoT) devices scale to permeate the Internet,
security concerns will become increasingly important as IoT devices
can be easily compromised if software updates are not done.
An example of a security attack is denial-of-service (DoS) attacks.
One promising application for blockchain technology is doing resource provisioning over the Internet.
This can come in the form of provisioning network resources for quality-of-service (QoS) traffic
or provisioning against over usage of networks that lead to DoS attacks.
Given that network resources are shared across multiple entities,
one can use blockchain as a tamper-evident logging mechanism to track and enforce
network resource utilization of edge devices and service providers in a hierarchical manner.

\noindent
{\bf Secure network slicing.}
Within the core of the 5G network,
network services can be broken into several private slices, one for each tenant.
Network services are deployed as virtualized network functions that are
replicated across nodes for fault tolerance.
These network functions can leverage blockchains to provide
tamper-evident communication and cross-domain transactions in the cloud~\cite{zhou2011secure}.
\section{System Model}\label{sec:model}

In a blockchain system, nodes agree on their shared states
across a large network of possibly {\em untrusted} participants.
While in a permissionless blockchain, e.g., Bitcoin \cite{nakamoto2008bitcoin},
the network is public, and anyone can participate without a specific identity,
a {\em permissioned} blockchain system,
e.g., Hyperledger Fabric \cite{androulaki2018hyperledger},
consists of a set of known, identified but possibly untrusted nodes.
\sys is a permissioned blockchain system consisting of a distributed set of 
edge devices, edge servers, fog servers, and cloud servers,
organized in a hierarchical tree structure.
Each logical vertex of the tree, called {\em a domain}, consists of a number of nodes sufficient to guarantee {\em fault tolerance}
(except for height$-$0 domains where the number of edge devices might not be known).

Nodes within each domain follow either the crash or the Byzantine failure model.
In the crash failure model, nodes may fail by stopping, and may restart,
whereas, in the Byzantine failure model,
faulty nodes may exhibit arbitrary and potentially malicious behavior.
Crash fault-tolerant (CFT) protocols, e.g., Paxos \cite{lamport2001paxos},
guarantee safety in an asynchronous network using $2f{+}1$ crash-only nodes
to overcome $f$ simultaneous crash failures while
in Byzantine fault-tolerant (BFT) protocols, e.g., PBFT \cite{castro1999practical},
$3f{+}1$ nodes are usually needed to guarantee safety
in the presence of $f$ malicious nodes \cite{bracha1985asynchronous}.

\begin{figure}[t] \center
\includegraphics[width=0.6\linewidth]{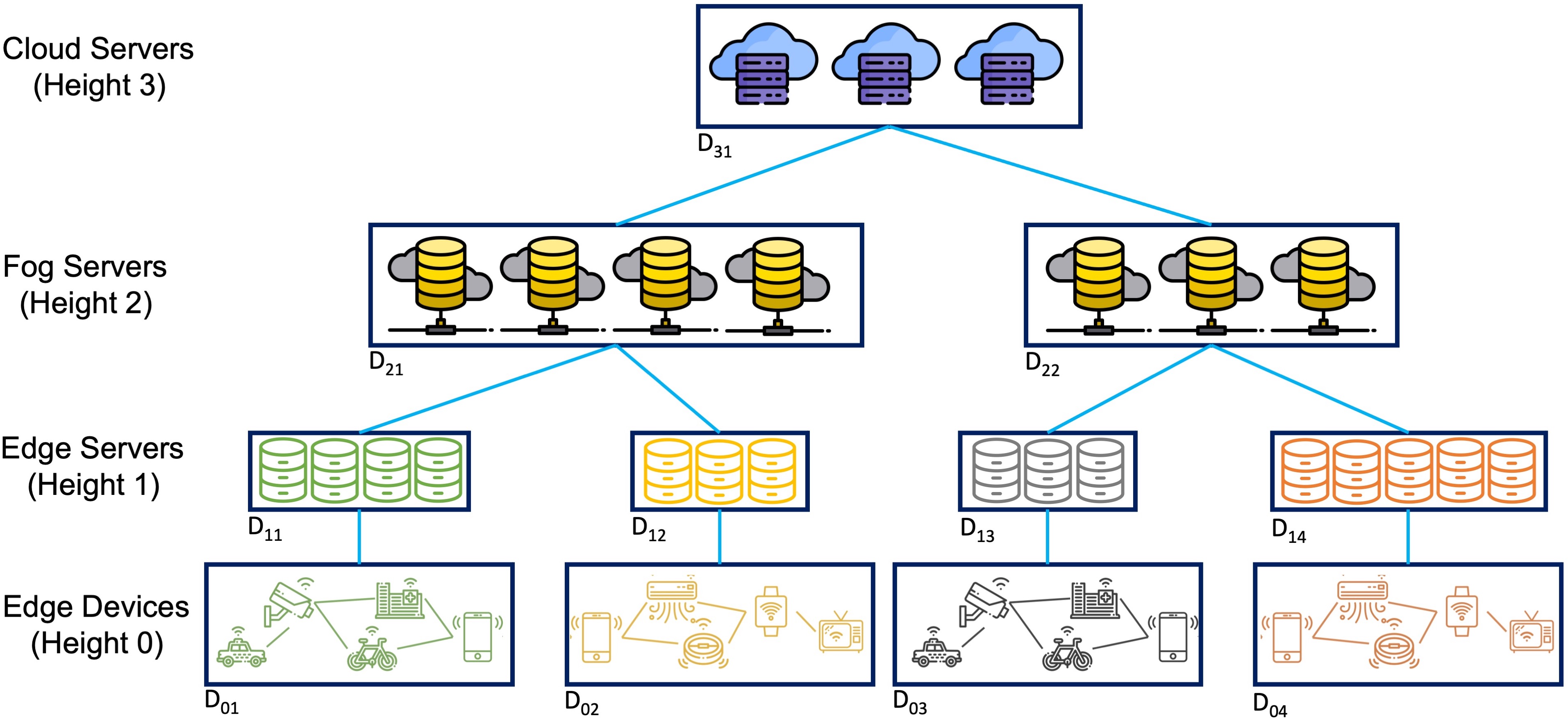}
\caption{The \sys deployment on an edge commuting network}
\label{fig:edge}
\end{figure}

Figure~\ref{fig:edge} presents a sample $4$-layer \sys deployment on an edge network
consisting of $11$ domains.
For example, $D_{21}$ includes $4$ nodes that follow Byzantine failure model ($3f+1$ nodes where $f=1$) while
$D_{14}$ consists of $5$ nodes that follow crash failure model ($2f+1$ nodes where $f=2$).

\ifextend{Each (height$-$1 and above) node is aware of its path to the root of the hierarchy at the domain level
-- a reasonable assumption since these domains loosely maps into public Internet-service-provider domains.
The network infrastructure and domains are reconfigurable, e.g.,
new servers can be added as long as the domains are still fault-tolerant.
If the underlying network infrastructure is reconfigured, ancestor/descendant domains will be informed.
When an edge device wants to join a leaf domain, it is authenticated by the corresponding parent height$-$1 domain using
a unique non-forgeable device id. The id is used for all future communication.}\fi

\sys assumes the partially synchronous communication model as it is typically used in practical fault-tolerant protocols.
In the partial synchrony model, 
an unknown global stabilization time (GST) exists, after which
all messages between correct replicas are received within some unknown bound.
\sys further inherits the standard assumptions of existing fault-tolerant systems, including
the unreliability of the network,
the existence of point-to-point bi-directional communication channels to connect nodes, and
a strong adversary that can coordinate malicious nodes but cannot subvert standard cryptographic assumptions.
\sys also uses digital signatures and public-key infrastructure (PKI).
We denote a message $m$ signed by node $r$ as
$\langle m \rangle_{\sigma_r}$ and
the digest of a message $m$ by $\Delta(m)$.
\ifextend{For signature verification, we assume that
nodes have access to the public keys of the required nodes, e.g., nodes on its path to the root.}\fi

The main underlying data structure in blockchain systems is the {\em blockchain ledger},
an append-only replicated structure that is shared among participants.
\sys follows the edge computing paradigm and brings computation and data closer to the network edges where
height$-$1 domains execute transactions.
\ifextend{Each height$-$1 domain executes a separate set of transactions.
In order to reduce communication overhead across domains over the wide-area network,
each height$-$1 domain maintains its own ledger where
the ledger is replicated on all nodes of the domain to provide fault tolerance.}\fi

\sys targets edge computing applications where data accesses have an affinity towards locality.
As a result, in \sys, each height$-$1 domain maintains its own ledger
replicated on all nodes of the domain to provide fault tolerance.
This design choice demonstrates a trade-off between performance and availability.
On one hand, replicating data on a single domain leads to high performance because
\sys does not need to deal with costly cross-domain replication protocols for every transaction.
On the other hand, the availability of \sys is reduced in case an entire domain fails, e.g., due to natural disasters like tornadoes or earthquakes.
This is in contrast to geo-replicated systems \cite{agrawal2013managing}\cite{baker2011megastore}\cite{corbett2013spanner}\cite{nawab2013message}\cite{nawab2015minimizing}
where data is replicated on all domains (clusters), and the system is able
to tolerate the failure of an entire domain.

Edge devices send their transaction requests
to their height-$1$ parent domains, e.g.,
leaf domain $D_{02}$ is connected to a height$-$1 domain $D_{12}$.
In a height-$1$ domain, due to data dependency among transactions of the same domain,
transactions are {\em totally ordered} to ensure data consistency.
The total order of transaction blocks in the blockchain ledger is captured by {\em chaining} blocks together, i.e.,
each block includes the cryptographic hash of the previous block.

In addition to the blockchain ledger, edge servers maintain the {\em blockchain state}.
The blockchain state is a datastore that maintains data and is being updated by executing transactions.
Similar to the blockchain ledger, each domain's blockchain state is replicated on the nodes of the domain.

\ifextend
Transactions are initially initiated by edge devices and processed in order by edge severs ({\em height$-$1} domains).
However, depending on the infrastructure and due to the lack of connectivity between edge devices and edge servers or
to offload traffic from edge servers, consensus among edge devices within a leaf domain 
and independent of edge servers in order to order transactions might also be needed.
This is also consistent with the D2D feature of 5G networks.
\fi

\sys uses the hierarchical structure of edge networks to provide four main functionalities.
First, \sys processes cross-domain transactions using a coordinator-based approach
by relying on the lowest common ancestor of all involved domains, resulting in lower latency (Section~\ref{sec:cons}).
Second, \sys enables data aggregation by propagating (a summarized version of) the ledgers up the hierarchy (Section~\ref{sec:propagation}).
Third, height$-$1 domains can optimistically process cross-domain transactions independent of each other
and rely on higher-level nodes to detect inconsistencies (Section~\ref{sec:opt}).
Finally, \sys supports the mobility of nodes by relying on edge servers in the local and remote height$-$1 domains (Section~\ref{sec:mobile}).

\section{Coordinator-based Consensus Protocol}\label{sec:cons}

Processing transactions requires establishing consensus on a unique order of client requests.
In \sys, transactions are initiated by edge devices (height$-$0) and executed by edge servers in height$-$1 domains.
\ifextend
Consensus protocols use the State Machine Replication (SMR) algorithm \cite{lamport1978time}.
The algorithm has to satisfy four main properties \cite{cachin2011introduction}:
(1) {\em agreement:} all non-faulty nodes must agree on the same value,
(2) {\em Validity (integrity):} a committed value must have been proposed by some (non-faulty) node,
(3) {\em Consistency (total order):} all non-faulty nodes commit the same values in the same order, and
(4) {\em termination:} eventually every node commits some value.
The first three properties are known as {\em safety} and termination is known as {\em liveness}.
\sys guarantees safety in an asynchronous network, however, a synchrony assumption is required to provide liveness
(due to FLP impossibility result\cite{fischer1985impossibility})
\fi
Transactions are either internal, i.e., access records within a single domain, or
cross-domain, i.e., access records across different height$-$1 domains.

The internal consensus protocol is needed among the nodes within a single domain.
Edge servers within a height$-$1 domain
establish consensus on every \req received from edge devices (i.e., clients).
The \req messages are sent by edge devices to the {\em primary} (a pre-elected node that initiates consensus)
of the corresponding height$-$1 domain.
Based on the failure model of nodes,
\sys uses
a CFT protocol, e.g., Paxos~\cite{lamport2001paxos},  or a BFT protocol, e.g., PBFT~\cite{castro1999practical}.

Cross-domain transactions access records across different height$-$1 domains,
e.g., a micropayment transaction where the sender and recipient belong to two different domains.
To ensure data consistency, such transactions are appended to
the ledgers of all involved domains in the same order.
The coordinator-based approach in \sys is inspired by
the traditional coordinator-based commitment protocols in distributed databases.
However, \sys leverages the hierarchical structure of edge networks by relying on the Lowest Common Ancestor (LCA) domain of all involved height$-$1 domains (participants)
to play the coordinator role.
Since the hierarchy is structured based on the geographical distance of nodes,
the LCA domain has the optimal location to minimize the total distance (i.e., latency).
In comparison to existing coordinator-based approaches, \sys deals with several new challenges.

First, in \sys, in contrast to distributed databases where all nodes follow the crash failure model,
the coordinator and the involved domains (participants)
might follow different failure models.
As a result, messages from a Byzantine domain
must be certified by at least $2f+1$ (out of $3f+1$) nodes of the domain
(since the primary node might be malicious).

Second, in contrast to the coordinator-based approaches where a single coordinator (node or domain)
sequentially orders all cross-domain transactions, in \sys, there are multiple independent coordinator domains in the network, i.e.,
any domains in height$-$2 and above could be a coordinator (an LCA domain).
As a result, a participant domain in addition to its internal transactions,
might be involved in several concurrent independent cross-domain transactions
ordered by separate coordinator domains at the same time.

Finally, while \sys processes cross-domain transactions in parallel,
ensuring consistency between concurrent order-dependent transactions is challenging especially when
the read-set and write-set of transactions are unknown beforehand, hence,
existing techniques \cite{thomson2012calvin}\cite{faleiro2017high}\cite{faleiro2015rethinking} can not be used.
\ifextend
For example, if two concurrent cross-domain transactions
$m$ (between $d_i$, $d_j$, and $d_k$) and
$m'$ (between $d_i$, $d_j$, and $d_l$) are initiated,
\sys must guarantee that $m$ and $m'$ are appended
to the ledger of both $d_i$ and $d_j$ domains in the same order,
i.e., either $m {\rightarrow} m'$ or $m' {\rightarrow} m$.
Since the LCA of $d_i$, $d_j$, and $d_k$ might be different from the LCA of $d_i$, $d_j$, and $d_l$,
\sys can not rely on the LCA domain to guarantee consistency.
\fi

\subsection{Coordinator-based Cross-Domain Protocol}

The normal case operation of the coordinator-based protocol is presented in Algorithm~\ref{alg:cross-coordinator}.
Although not explicitly mentioned, every sent and received message is logged by nodes.
As indicated in lines 1 to 5,
$d_c$ is the coordinator domain,
$\pi(d)$ represents the primary node of domain $d$,
$D$ is the set of involved domains in the transaction,
$\pi(D) = \{\pi(d) | d \in D\}$ is the set of primary nodes of the involved domains.

\noindent
{\bf Prepare phase.}
Once the primary node of an involved domain receives a valid cross-domain transaction $m$,
as shown in lines $6{-}7$, the primary node forwards it {\em directly} to all nodes of the LCA domain $d_c$ of the involved domains.
Upon receiving a cross-domain transaction (lines $8{-}11$),
the primary of the LCA domain, $\pi(d_c)$, validates the message.
Since \sys assumes that the read-set and write-set of transactions are unknown beforehand,
fine-grained locking mechanisms that lock the accessed records do not work.
As a result, if the primary node $\pi(d_c)$ is currently processing another cross-domain transaction $m'$
(i.e., has not sent \three message for $m'$)
where
the involved domains of two requests $m$ and $m'$ intersect in at least two domains,
the node does not process the new request $m$ before the earlier request $m'$ gets committed.
This is needed to ensure consistency, i.e., cross-domain requests are committed in the same order on overlapping domains.
Otherwise, node $\pi(d_c)$ assigns a sequence number $n_c$ to $m$ and initiates
consensus on request $m$ in the coordinator domain $d_c$.
Once consensus is established, the primary node $\pi(d_c)$
sends a signed \pre message
including the sequence number $n_c$, request $m$ and its digest $\delta=\Delta(m)$
to the nodes of all involved domains.
Note that if the nodes of the LCA domain follow the Byzantine failure model,
a {\em certificate} consisting of $2f{+}1$ signed (\three) messages is needed.

\newcommand{\prepc}{{\tiny $\langle\text{\PRE}, n_c, \delta, m \rangle_{\sigma}$}\xspace}
\newcommand{\predc}{{\tiny $\langle\text{\PRED}, n_c, n_i, \delta, r \rangle_{\sigma}$}\xspace}
\newcommand{\threec}{{\tiny $\langle\text{\THREE}, n_i{-}n_j{-}{...}{-}n_k, \delta, r \rangle_{\sigma}$}\xspace}
\newcommand{\ackc}{{\tiny $\langle\text{\ACK}, n_c, n_i{-}n_j{-}{...}{-}n_k, \delta, r \rangle_{\sigma_r}$}\xspace}

\begin{algorithm}[t]
\caption{{\small Coordinator-based Cross-Domain Consensus}}
\label{alg:cross-coordinator}
\begin{algorithmic}[1]
\State {\em init():} 
\State \quad $r$ := {\em node\_id}
\State \quad $d_c$ := coordinator (lowest common ancestor) domain
\State \quad $\pi(d)$ := the primary node of domain $d$
\State \quad $\pi(D) = \{\pi(d) | d \in D\}$
\State upon receiving request $m$ and $r \in \pi(D)$
\State \quad forward request $m$ to $d_c$
\State upon receiving request $m$ and $r$ is $\pi(d_c)$
\State \quad if $r$ is not processing $m'$
where $m$ and $m'$ intersect
\State \qquad establish consensus on $m$ among nodes in $d_c$
\State \qquad send signed \prepc to all domains $D$ 
\State upon receiving \prepc message(s) and $(r=\pi(d_i) \in \pi(D))$
\State \quad if $r$ is not processing request $m'$
where $m$ and $m'$ intersect
\State \qquad establish consensus on the message among nodes in $d_i$
\State \qquad send signed \predc to $d_c$
\State upon receiving \predc from $D$ and $r == \pi(d_c)$
\State \quad establish consensus on the order of $m$ within $d_c$
\State \quad multicast signed \threec to all domains $D$
\State upon receiving \threec message and $r \in D$
\State \quad append the transaction and the {\scriptsize \sf commit} message to the ledger
\State \quad send \ackc to $\pi(d_c)$
\end{algorithmic}
\end{algorithm}

\noindent
{\bf Prepared phase.}
Upon receiving a valid \pre message,
as shown in lines $12{-}15$,
if the primary $\pi(d_i)$ of an involved domain $d_i$ is {\em not} processing
another cross-domain transaction $m'$ where
the involved domains of two requests $m$ and $m'$ intersect in at least two domains,
the primary $\pi(d_i)$ assigns a sequence number $n_i$ to $m$ and
initiates consensus in $d_i$ on its order.
Once consensus is achieved,
the primary $\pi(d_i)$ of each involved domain $d_i$
sends a signed (certified) \pred message to nodes of $d_c$
including both sequence numbers $n_c$ and $n_i$, request digest $\delta$, and node id $r=\pi(d_i)$.

\noindent
{\bf Commit phase.}
When primary node $\pi(d_c)$ of the coordinator domain receives valid \pred messages from every involved domain (lines $16{-}18$),
it establishes consensus within the coordinator domain and sends a certified \three message including a sequence number $n_i -n_j - ... n_k$
(i.e., concatenation of the received sequence numbers from
all involved domains) and request digest $\delta$
to every node of all involved domains.
Otherwise (if some involved domain has not agreed with the transaction), the domain sends a signed \abort message.

\noindent
{\bf Execution phase.}
Upon receiving a valid \three message (lines $19{-}21$),
each node considers the transaction as committed and
sends an \ack message to the coordinator domain.
If all transactions with lower sequence numbers have been executed, the node executes the transaction.
This ensures that all nodes execute transactions
in the same order as required to ensure safety.
Depending on the application, a \reply message including the execution results
might also be sent to the edge device (requester)
by either the primary (if nodes are crash only) or all nodes (if nodes follow Byzantine failure)
of the domain that has received the request.

\begin{figure}[t] \center
\includegraphics[width=0.6\linewidth]{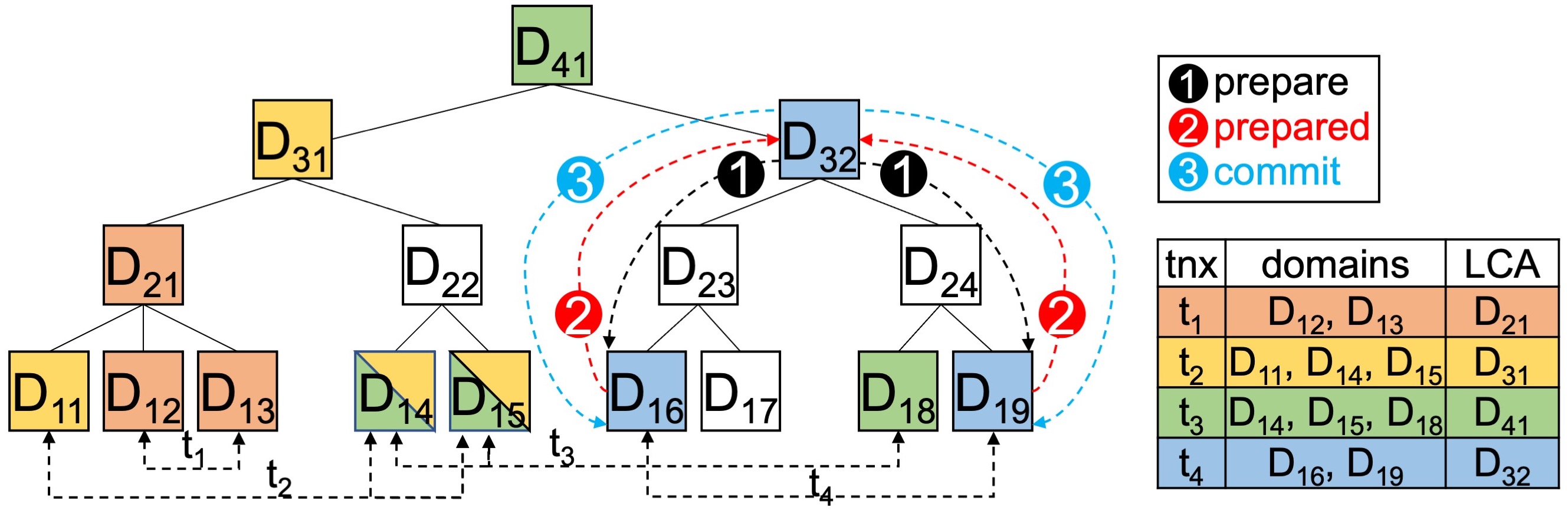}
\vspace{-1em}
\caption{Coordinator-based Cross-Domain Consensus}
\label{fig:coordinator}
\end{figure}

Figure~\ref{fig:coordinator} presents four different cross-domain transactions $t_1$ to $t_4$, their involved domains and
the LCA domain for each transaction, e.g.,
$D_{21}$ is the LCA domain of transaction $t_1$ between $D_{12}$ and $D_{13}$.
$D_{31}$ is the LCA domain of transaction $t_2$ between $D_{11}$, $D_{14}$ and $D_{15}$,
$D_{41}$ is the LCA domain of transaction $t_3$ between $D_{14}$, $D_{15}$ and $D_{18}$, and
$D_{32}$ is the LCA domain of transaction $t_4$ between $D_{16}$ and $D_{19}$.
To process transaction $t_4$,
\pre, \pred, and \three messages are directly exchanged between participants ($D_{16}$ and $D_{19}$) and their
LCA domain $D_{32}$ without the participation of the domains on the paths from participants to the LCA domain, e.g., $D_{23}$ and $D_{24}$.

In a situation
where cross-domain transactions
(1) are concurrent,
(2) overlap on at least two domains,
(3) are processed by different LCA domains, and
(4) their \pre messages are received by overlapping domains in a different order,
ensuring consistency might result in a deadlock situation.
This is because domains do not process the later transaction before receiving the \three message of the earlier transaction
(Algorithm~\ref{alg:cross-coordinator}, line $13$) to ensure consistency in a coarse-grained manner.
Note that if the LCA of transactions is the same,
the LCA does not initiate the second transaction and deadlock will not occur (Algorithm~\ref{alg:cross-coordinator}, lines $9{-}10$).
\ifextend
For example, in Figure~\ref{fig:coordinator}, if the \pre messages for two concurrent transactions
$t_2$ between $D_{11}$, $D_{14}$, and $D_{15}$ (sent by LCA domain $D_{31}$) and
$t_3$ between $D_{14}$, $D_{15}$, and $D_{18}$ (sent by LCA domain $D_{41}$)
are received by $D_{14}$ and $D_{15}$ in a different order, i.e., on one $t_2 {\rightarrow} t_3$ and on the other $t_3 {\rightarrow} t_2$,
deadlock happens.\fi

To resolve the deadlock, once the timer of an LCA domain for its cross-domain transaction is expired,
the LCA aborts the transaction and sends a new \pre message to the involved domains.
\sys assigns different timers to different domains to prevent consecutive deadlock situations.

\subsection{Primary Failure Handling}\label{sec:viewchange}

If the primary of either the LCA domain or a participant domain is faulty,
the primary failure handling routine of the internal consensus protocol,
e.g., view change in PBFT \cite{castro1999practical}, is triggered by timeouts
to elect a new primary.

For cross-domain transactions,
if node $r$ of an involved domain does not receive
a \three message from the LCA domain for a prepared request and its timer expires,
the node sends a
$\langle\text{\scriptsize \CMTQ}, n_c, n_i, \delta, r \rangle_{\sigma_r}$ message
to all nodes of the LCA domain where $n_c$ and $n_i$ are the sequence numbers assigned by
the primary nodes of LCA and $d_i$ domains and $\delta$ is the digest of the request.
Similarly, if node $r$ in the LCA domain has not received \pred message from an involved domain soon enough,
it sends a $\langle\text{\scriptsize \PREDQ}, n_c, \delta, r \rangle_{\sigma_r}$
to all nodes of the involved domain.

In either case, if the message has already been processed, the nodes simply re-send the corresponding response.
Nodes also log the query messages to detect denial-of-service attacks initiated by malicious nodes.
If the query message is received from $n-f$ nodes of a domain,
the primary will be suspected to be faulty resulting in running the failure handling routine.

Note that since in all communications between a participant and an LCA domain,
the primary of the sender domain multicasts messages, e.g., \req, \pre, or \pred,
to all nodes of the recipient domain,
if the primary of the recipient domain does not initiate consensus on the message in its domain
(even after other nodes relay the message to the primary),
it will eventually be suspected to be faulty.

Finally, if an edge device does not receive \reply soon enough, it multicasts the request
to all nodes of the domain that it has sent its request.
If the request has already been processed,
the nodes simply send the result back to the edge device.
Otherwise, if the node is not the primary, it relays the request to the primary.
If nodes do not receive \pre messages, the primary will be suspected to be faulty,
i.e., it has not multicast request to the LCA domain.

\subsection{Correctness}\label{sec:correct-coord}

We briefly analyze the safety (agreement, validity and consistency) and
the liveness of the coordinator-based protocol.

\begin{lmm} (\textit{Agreement})
If node $r$ commits request $m$ with sequence number $h$,
no other non-faulty node commits request $m'$ ($m \neq m'$) with the same sequence number $h$.
\end{lmm}

\begin{prf}
We assume that the internal consensus protocol of all domains
ensures agreement.
Let $m$ and $m'$ ($m \neq m'$) be two committed cross-domain requests with sequence numbers
$h = [h_i,h_j,h_k,...]$ and $h' = [h'_k,h'_l,h'_m,..]$ respectively.
Committing a request requires
matching \pred messages from $n{-}f$ different nodes of {\em every} involved domain.
Therefore, given an involved domain $d_k$ in the intersection of $m$ and $m'$,
at least a quorum of $n{-}f$ nodes of $d_k$ have sent matching \pred messages for $m$ and
at least a quorum of $n{-}f$ nodes of $d_k$ have sent matching \pred messages for $m'$.
Since any two quorums intersect on at least one non-faulty node, $h_k \neq h'_k$, hence, $h \neq h'$.
\end{prf}

\begin{lmm} (\textit{Validity})
If a non-faulty node $r$ commits $m$, then $m$ must have been proposed by some node $\pi$.
\end{lmm}

\begin{prf}
If nodes are crash-only, validity is ensured since crash-only nodes do not send fictitious messages.
With Byzantine nodes, validity is guaranteed based on
standard cryptographic assumptions
which the adversary cannot subvert (as explained in Section~\ref{sec:model}).
Since all messages are signed (by $2f+1$ nodes) and 
the request or its digest is included in each message
(to prevent changes and alterations to any part of the message),
if request $m$ is committed by non-faulty node $r$, the same request must have been proposed earlier
by some node $\pi$.
\end{prf}

\begin{lmm} (\textit{Consistency})
Let $D_\mu$ denote the set of involved domains (participants) for a request $\mu$.
For any two committed requests $m$ and $m'$ and any two nodes $r_1$ and $r_2$
such that $r_1 \in d_i$, $r_2 \in d_j$, and $\{d_i,d_j\} \in D_m \cap D_{m'}$,
if $m$ is committed before $m'$ in $r_1$, then $m$ is committed before $m'$ in $r_2$.
\end{lmm}

\begin{prf}
As shown in lines $12{-}15$ of Algorithm~\ref{alg:cross-coordinator},
when node $r_1$ of a participant domain $d_i$ receives a \pre message for some cross-domain transaction $m$,
if the node is involved in another uncommitted cross-domain transaction $m'$
where some other domain $d_j$ is also involved in both transactions,
node $r_1$ does not send a \pred message for transaction $m$ before $m'$ gets committed.
Since committing request $m$ requires a quorum of \pred messages from {\em every} involved domains,
$m$ cannot be committed until $m'$ is committed.
As a result, the order of committing messages is the same in all involved domains.
The coordinator domain $d_c$ also checks the same condition before sending \pre messages (lines $8{-}11$).
\end{prf}

\begin{lmm}(\textit{Liveness})
A request $m$ issued by a correct client eventually completes.
\end{lmm}

\begin{prf}
Due to the FLP result \cite{fischer1985impossibility}, \sys
guarantees liveness {\em only} during periods of synchrony.
\sys addresses liveness in primary failure and deadlock situations.
First, if the primary of a domain is faulty, e.g.,
does not multicast valid \req, \pre, \pred, or \three messages, as explained earlier,
its failure will be detected and using
the primary failure handling routine of the internal consensus protocol,
a new primary will be elected.
Second, \sys addresses deadlock situations resulting from
concurrent cross-domain transactions that are
received by overlapping domains in different orders.
\end{prf}
\section{Lazy Propagation of Blockchain Ledgers}\label{sec:propagation}

\sys enables height$-$2 and above domains to perform data aggregation over transactions executed by edge servers in height$-$1.
To this end, such domains need to maintain (a summarized version of) the ledgers of their child domains.

To send transaction blocks up the hierarchy, edge servers proceed through a succession of {\em rounds}.
Each round ends after some predefined time interval that is identical for all height$-$1 domains.
At the end of each round $r_n$, each height$-$1 domain
sends a \textit{\block} message to its parent domain.
The \block message includes
(1) all transactions that are appended to the ledger in that round,
(2) the Merkle hash tree of those transactions used to verify the content of the block, and
(3) an application-dependent abstract version of the blockchain state updates in that round, i.e., $\lambda (D^{r_n}-D^{r_{n-1}})$
where $D^{r_n}$ and $D^{r_{n-1}}$ are the blockchain states at the end of rounds $r_n$ and $r_{n-1}$ and 
the abstraction function $\lambda$ is deterministic, predefined, and known by all nodes.
For example, in a ridesharing application,
it might be sufficient to send only the working hour attribute of the records that are updated
to the higher-level nodes.
If a domain has not received any transaction in that round, it sends an empty \block message.

Depending on the failure model of the child domain,
the \block message is signed (certified) by
either the primary (in the crash failure model) or
at least $2f+1$ nodes (in the Byzantine failure model), i.e.,
the primary constructs a {\em certificate} consisting of $2f+1$ \three messages proving that
consensus has been achieved on the \block message within the child domain.
Threshold signature can also be used to replace $2f+1$ signatures with a single threshold signature \cite{shoup2000practical}\cite{cachin2005random}.

Nodes in higher-level domains, on the other hand, achieve (internal) consensus on \block messages
that they receive from child domains. 
The \block messages are sent by the primary node of a domain to all nodes of its parent domain.
These \block messages contain a collection of committed transactions in the most recent time interval.
Broadcasting of \block messages to all nodes in the parent domain enables nodes of the parent domain to detect malicious behavior of primary nodes.

If the primary node of the parent domain
has not received the \block message from a child domain
after a predefined time (e.g., the primary of the child domain might be faulty),
it sends a query message to all nodes of the child domain.
To ensure that the completion of each round is deterministic on all nodes of a domain,
the primary node puts a {\sf \small "cut"} sign into the \one message of the last request
informing other nodes of the completion of a round.
Since nodes establish consensus on received messages, if a primary
sends the {\sf \small "cut"} sign maliciously, 
it will be easily detected.

\begin{figure}[t] \center
\includegraphics[width=0.6\linewidth]{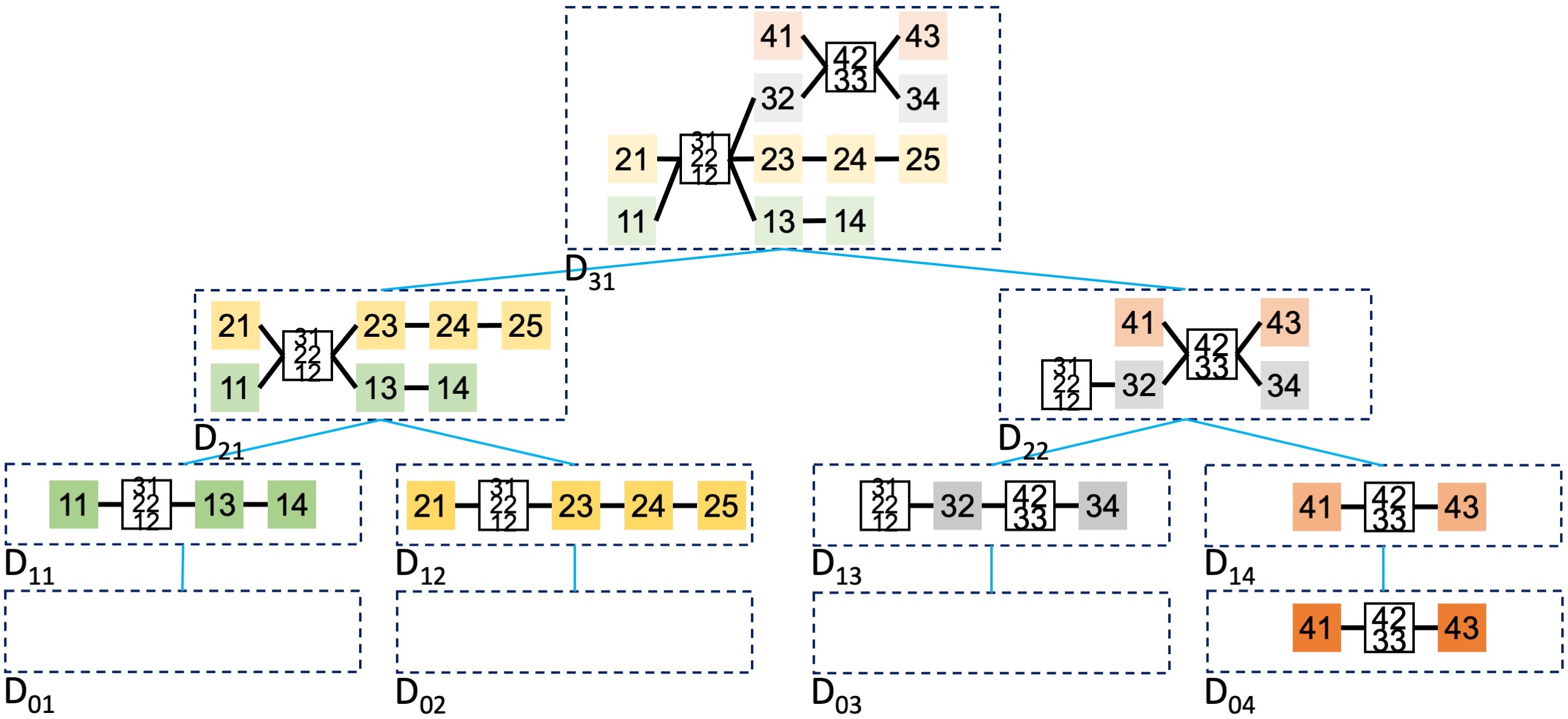}
\caption{An Example of \sys Blockchain Ledger}
\label{fig:ledger}
\end{figure}

Since height$-$2 and above domains might have multiple child nodes, each domain receives \block messages from possibly multiple child domains and
orders all transactions within received messages at each round.
If there is no dependency between transactions of different child domains,
any unique order of transactions is acceptable.
However, cross-domain transactions,
which are appended to the ledger of multiple child domains,
must be appended to the ledger of the parent domain only once.
Therefore, the resulting ledger
is a {\em directed acyclic graph} to capture the order dependencies.

Similar, to height$-$1 domains, each higher-level domain sends
\block messages (with the same structure) to its parent domain at regular predefined time intervals in a lazy fashion.
The time interval for the domains at the same level is the same, however,
domains at higher levels may have larger time intervals to reduce communication overhead.
Finally, the ledger of the root domain consists of all transactions that are processed in the system, and
its blockchain state is a summarized view of all blockchain states in the network.

Figure~\ref{fig:ledger} presents a set of transactions and shows how these transactions are appended to the blockchain
ledger of different domains at different heights for the same network as Figure~\ref{fig:edge}
(leaf domains are not shown because they do not maintain blockchain ledgers).
The presented ledger of each domain is replicated on all nodes of that domain.
In the figure, one block denotes one transaction. The sequence number of each transaction presents the order of
the transaction within the ledger.
While each internal transaction of a domain has a single-part sequence number, e.g., $11$, $13$, and $14$,
cross-domain transactions have multi-part sequence numbers where each part demonstrates the order of
the transaction in an involved domain, e.g., $12{-}22{-}31$, is a cross-domain transaction among $D_{11}$, $D_{12}$, and $D_{13}$.
\ifextend{As can be seen, the ledgers of height$-$1 domains are linear chains.}\fi

\begin{figure}[t] \center
\includegraphics[width=0.6\linewidth]{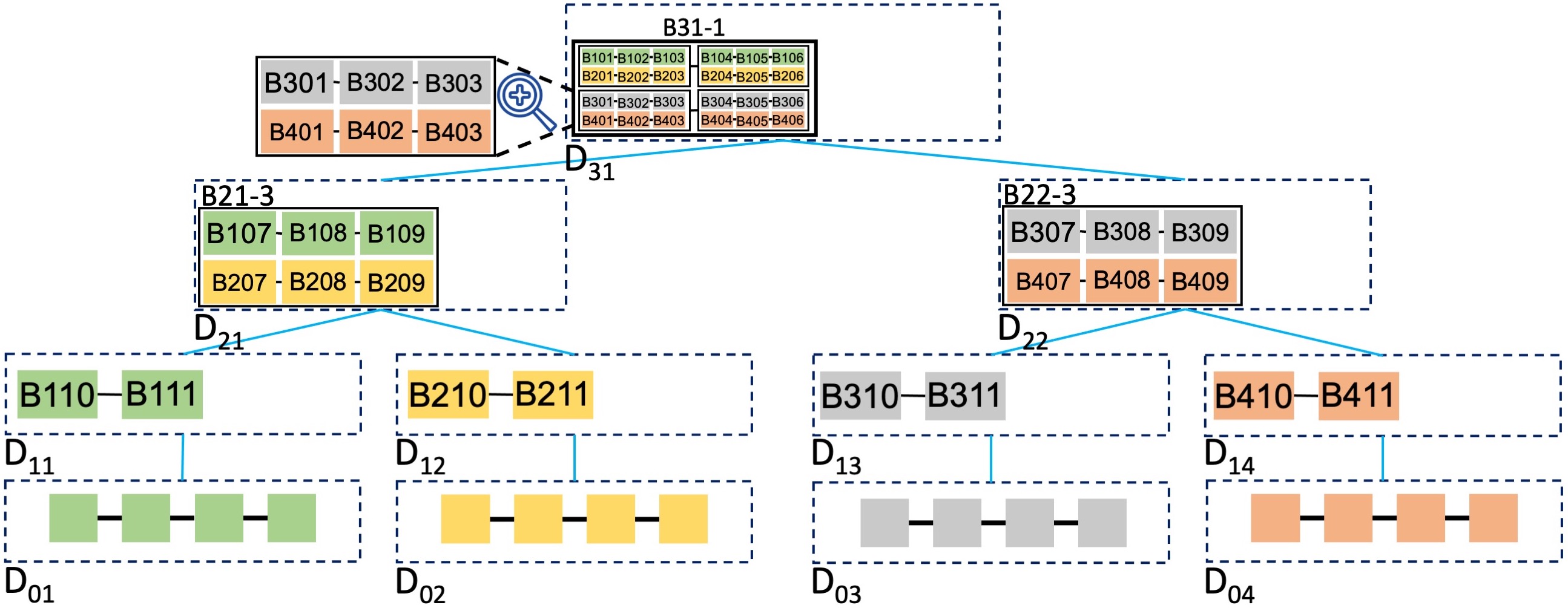}
\caption{A Snapshot of \sys}
\label{fig:snapshot}
\end{figure}

Figure~\ref{fig:snapshot} presents a snapshot of \sys for the network of Figure~\ref{fig:edge}.
This snapshot shows the lazy propagation of blockchain updates via the hierarchy.
Each height$-$1 domain $D_x$ in its $n$-th round appends transactions to its ledger to construct block $Bx{-}n$, e.g.,
height$-$1 domain $D_{11}$ is in its $6$th round constructing $B11{-}06$.
The height$-$2 domain $D_{22}$ is in its third round constructing block $B22{-}03$.
Domain $D_{22}$ has received $B13{-}05$ and $B14{-}05$ from its child domains $D_{13}$ and $D_{14}$ in this round.
Finally, the root domain $D_{31}$ has appended transaction blocks
$B21{-}01$ and $B21{-}02$ (received from $D_{21}$) and $B22{-}01$, and $B22{-}02$ (received from $D_{22}$)
to its ledger where, for example, block $B21{-}01$ itself contains four transaction blocks
$B11{-}01$, $B11{-}02$, $B12{-}01$, and $B12{-}02$.
In this example, the time interval of height$-$2 domains is twice the height$-$1 domains.
Note that height$-$1 domains maintain their own blockchain states while
a summarized view of the
blockchain state is maintained by higher-level domains.
\section{Optimistic Consensus Protocol}\label{sec:opt}

\ifextend
The coordinator-based consensus protocol of \sys is more efficient than
the existing coordinator-based protocols because \sys (1) relies on the lowest common ancestor domain to minimize the distance between
the coordinator and involved (participant) domains, and (2) processes cross-domain transactions in parallel.
However, it still requires multiple rounds of communication between the coordinator and participant domains.
\fi
\sys leverages the lazy propagation of ledgers presented in Section~\ref{sec:propagation}
to enable the optimistic processing of cross-domain transactions.
In the optimistic protocol, each involved height$-$1 domain optimistically processes and commits a cross-domain transaction
independent of other involved domains, assuming
that all other involved domains also commit the transaction.
Since transactions will propagate up, nodes in higher levels and eventually
the LCA domain can check the commitment of the transaction.

In the optimistic approach, upon receiving a cross-domain request from an authorized edge device,
the primary of the initiator height$-$1 domain multicasts the request to all nodes of the involved height$-$1 domains.
The primary might behave maliciously by not sending the request to some involved domains.
Hence, upon receiving the request, all nodes of the initiator domain multicast the request to the involved domains ensuring that they all received the request.
Upon receiving a request, each involved domain (including the initiator domain),
uses its internal consensus protocol to optimistically
establish agreement on transaction order and executes it
(assuming all other involved domains also execute the transaction).

For each executed cross-domain transaction $t$, nodes of a domain maintain
a list of transactions (both internal and cross-domain) that are executed after $t$
and have direct or indirect data dependency to transaction $t$.
If transaction $t$ gets aborted, e.g., some other involved domain does not commit the transaction,
all data-dependent committed transactions need to be aborted as well.
The list is deleted once transaction $t$ has eventually been committed or aborted.

\begin{figure}[t] \center
\includegraphics[width=0.6\linewidth]{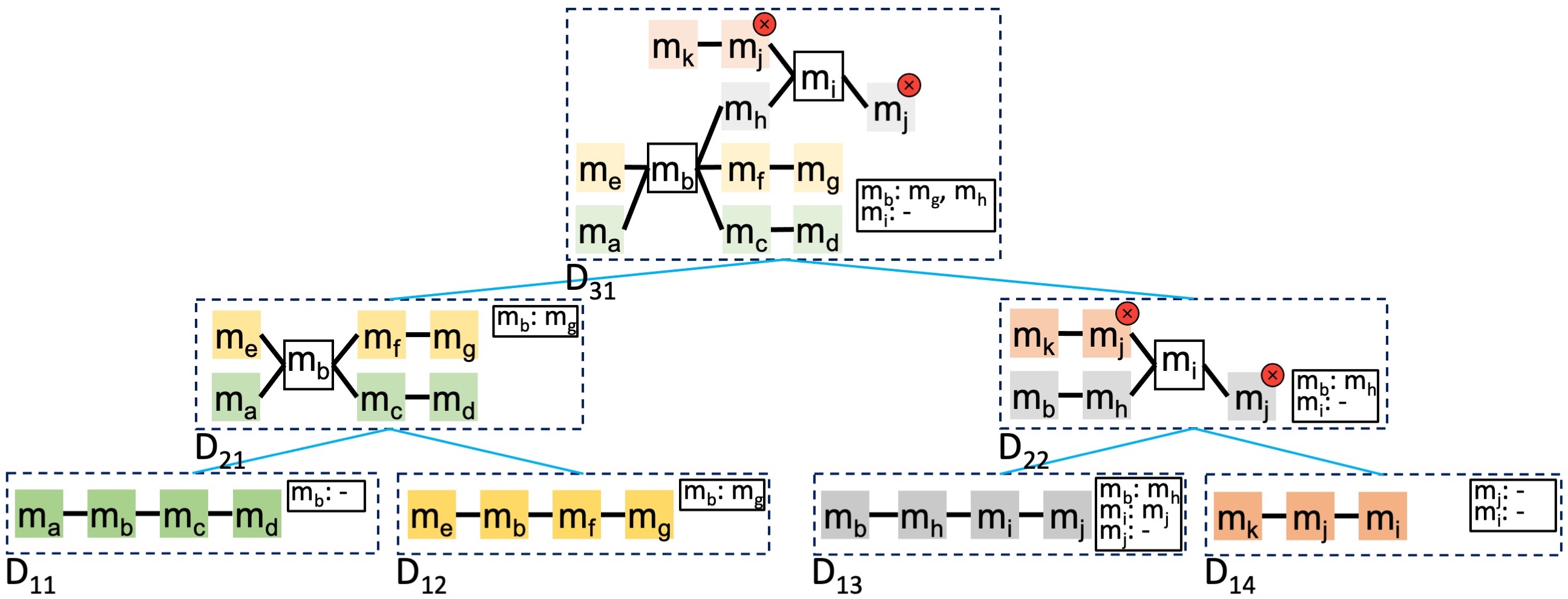}
\caption{Example of Optimistic Cross-Domain Consensus}
\label{fig:optimistic}
\end{figure}

Figure~\ref{fig:optimistic} presents the ledger of different domains using the optimistic cross-domain consensus protocol
for the same network as Figure~\ref{fig:edge}.
In this figure, $m_b$ is a cross-domain transaction between $D_{11}$, $D_{12}$, and $D_{13}$ and
$m_i$ and $m_j$ are between $D_{13}$, and $D_{14}$.
Each domain maintains a list of data-dependent transactions for each cross-domain transaction, e.g.,
in $D_{12}$, $m_g$ has data dependency to $m_b$. 

Each height$-$1 domain processes all internal and cross-domain requests and upon completion of a round, sends a \block message to its parent domain.
In the optimistic protocol, the \block message,
in addition to the committed transactions (blockchain ledger) and blockchain state,
consists of non-committed (aborted) cross-domain transactions (to inform other domains), and
the dependency lists for cross-domain transactions (within the current and previous blocks)
that have not yet been decided by all their involved domains.

Each parent domain and eventually the LCA of all involved domains in a cross-domain transaction
ensures that concurrent cross-domain transactions (if any)
have been appended to the ledger of the intersection domains in the same order.
Otherwise, (at least) one of the transactions will be aborted.
For example, in Figure~\ref{fig:optimistic}, $m_i$ and $m_j$ are appended
to the ledger of $D_{13}$ and $D_{14}$ in an inconsistent order, hence,
domain $D_{22}$ aborts (only) $m_j$.
\sys guarantees that aborting transactions is deterministic,
i.e., all higher-level domains reach the same decision on choosing transactions to abort, e.g.,
they all abort the transaction with the lowest id.
Note that intermediate domains between involved domains and the LCA domain might 
receive the transaction from a subset of involved domains and be able
to partially check the consistency and early abort in case of inconsistency.
For example, in Figure~\ref{fig:optimistic}, domain $D_{21}$ receives $m_b$ from $D_{11}$ and $D_{12}$ (but not $D_{13}$).

Upon finding an inconsistency, the primary of the domain marks the transaction and
all its data-dependent transactions as aborted, e.g., $m_j$ in Figure~\ref{fig:optimistic}.
The primary also sends a certified \abort message
including the request digest to the nodes of the involved domains.
Involved domains need to rollback the aborted transaction and its data-dependent ones.
% Since transactions have already been appended to the ledger, the domains execute
% compensating transactions to cancel their effect.

Each intermediate and eventually the LCA domain then checks 
whether the transaction is committed by the involved domains.
The intermediate domains can check the commitment of the transaction by a subset of the involved domains.
If the transaction is committed by all involved domains, the transaction will be appended to the ledger
and upon the completion of the round sent to the parent domain.
Once the primary of the LCA domain receives the transaction from all involved domains, it sends a signed \three message
to all domains informing them that the transaction is committed.

If the transaction has not been appended to the ledger (\block message) of an involved domain
(due to the asynchronous nature of the network),
the intermediate or the LCA domain does not append the transaction and waits for the next \block messages.
The domain also does not append the next transactions within the \block message to its ledger.
This is needed because there might be an inconsistency issue where the domain needs to mark the transactions as aborted.

\ifextend
If a node of an involved domain does not receive the \three message for cross-domain transaction $m$ after a predefined time,
it sends a signed \cmtq including the digest of $m$ to its parent domain.
If the parent domain has already been informed about the decision,
it sends the \three (or \abort) message to the node.
Otherwise, it forwards the \cmtq to its parent domain.
If the LCA domain receives the \cmtq and the transaction has been processed,
the domain sends the \three (or \abort) message to the node, otherwise,
it sends a \cmtq to the involved domain(s) that has not sent their transactions.
Upon receiving a valid \cmtq message,
if the domain has not processed the transaction, it immediately starts processing the transaction.
If the request is not committed in some predefined number of rounds by all involved domains,
it is considered to be aborted.
\fi

In the optimistic approach, 
the predefined time interval for completion of rounds (i.e., sending \block messages to the parent domains)
is smaller to detect inconsistencies in cross-domain transactions earlier.
This avoids too many cascaded aborts of transactions, as any inconsistencies will result in the abort of transactions that depend on the aborted transaction.

\medskip\noindent
{\bf Correctness.}
We now briefly show the safety and liveness of the optimistic approach.

\begin{lmm} (\textit{Agreement})
If node $r$ commits request $m$ with sequence number $h$,
no other correct node commits request $m'$ ($m \neq m'$) with the same sequence number $h$.
\end{lmm}

\begin{prf}
We assume the internal consensus protocols, e.g., Paxos and PBFT, guarantee agreement.
In the optimistic protocol,
the same cross-domain transaction has different sequence numbers in different domains, however,
it does not violate the agreement property, i.e.,
no two requests have the same sequence number in the same domain.
In addition, \sys prevents different domains to assign the same sequence number to different requests by
defining a prefix for the sequence numbers of each domain.
Moreover, if the transaction is not committed in a domain,
the LCA domain detects it resulting in aborting the transaction.
\end{prf}

\begin{lmm} (\textit{Validity})
If a correct node $r$ commits $m$, then $m$ must have been proposed by some correct node $\pi$.
\end{lmm}

\begin{prf}
Validity is guaranteed in the same way as coordinator-based cross-domain consensus (lemma 4.2).
\end{prf}

\begin{lmm} (\textit{Consistency})
Let $P_\mu$ denote the set of involved domains for a request $\mu$.
For any two committed requests $m$ and $m'$ and any two nodes $r_1$ and $r_2$
such that $r_1 \in p_i$, $r_2 \in p_j$, and $\{p_i,p_j\} \in P_m \cap P_{m'}$,
if $m$ is committed before $m'$ in $r_1$, then $m$ is committed before $m'$ in $r_2$.
\end{lmm}

\begin{prf}
As mentioned earlier, upon receiving a cross-domain transaction,
the LCA domain first checks the consistency.
Since $p_i$ and $p_j$ are involved in both $m$ and $m'$,
the LCA of both $m$ and $m'$ can detect any inconsistencies in the order of transactions in both domains
and resolve it by aborting either $m$ or $m'$.
The aborting strategy is deterministic and results in aborting the same transaction
on both LCAs, i.e.,
it does not matter which LCA receives the transactions first,
if there is an ordering inconsistency they both either abort $m$ or abort $m'$.
While transactions might be initially {\em optimistically} committed in an inconsistent order,
eventually inconsistency will be resolved, i.e., the protocol guarantees eventual consistency.
\end{prf}

\begin{prop}(\textit{Termination})
A request $m$ issued by a correct client eventually completes.
\end{prop}

The liveness of the algorithm is guaranteed in periods of synchrony
based on the assumption that LCA and involved domains
ensure liveness for all transactions.
\ifextend
Furthermore, if a node does not receive the \three (or \abort) message from the LCA domain
for some cross-domain request and its timer expires, as discussed earlier,
it sends a \cmtq message to the parent domain,
resulting in sending a \cmtq message to the nodes of the involved domain(s) that
have not committed the message.
\fi
If the request is not committed in some predefined number of rounds by all involved domains
it is considered to be aborted.

\ifextend
\subsection{Processing Transactions with Edge Devices}\label{sec:consedge}

In \sys, edge servers (height$-$1 domains) process transactions that are initiated by edge devices.
However, in the case of the lack of connectivity between edge devices and edge servers, e.g., in rural areas, or
to offload traffic from edge servers, \sys enables leaf (height$-$0) domains to process transactions among edge devices.
This is also consistent with the D2D feature of 5G networks.
In the case of lack of connectivity, assuming the domain is fault-tolerant,
edge devices within small leaf domains can establish consensus on the order of local transactions among themselves.
Batches of transactions can then be sent to the parent height$-$1 domain using \block messages
(with the same structure as before without the blockchain state).
The primary node of the parent domain will validate transactions within a \block message and initiate the internal consensus protocol
on the batch of transactions to be committed and executed by all nodes.

Leaf domains can also be seen as a layer-two solution \cite{gudgeon2020sok} to improve the performance of height$-$1 domains by
enabling small subsets of edge devices to perform transactions among themselves \cite{decker2015fast}.
In particular, inspired by off-chain lightning networks \cite{miller2019sprites,poon2016bitcoin,network2019raiden},
for asset-transfer applications, e.g., micropayment,
\sys can be extended to support channels between edge devices.
Channels are initiated by locking (some portion of) edge devices' assets on the height$-$1 blockchain state.
Edge devices then process transactions within the channel, and upon channel closure,
committed transactions will be sent to the height$-$1 domain.
\fi
\section{Mobile Consensus}\label{sec:mobile}

This section addresses the next challenge of \sys: processing transactions initiated by mobile edge devices. 
When an edge device moves from its {\em local} to a {\em remote} leaf domain,
reaching consensus on transactions that are initiated by the mobile device is challenging.
Specifically, since edge servers of the remote height$-$1 domain do not have access to the state of the mobile node, e.g.,
the account balance of the node in the micropayment application, they are not able to process its requests.
Moreover, any communication across domains goes through wide-area networks where bandwidth is more limited and subjected to higher latencies.

\begin{figure}[t] \center
\includegraphics[width=0.6\linewidth]{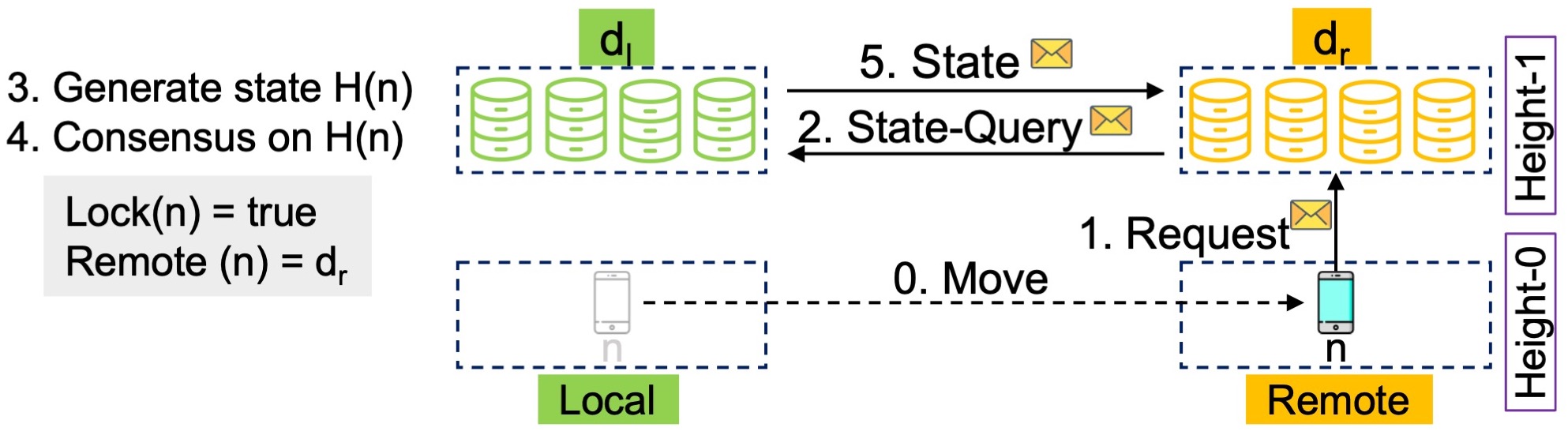}
\caption{Mobile Consensus}
\label{fig:mobilecons}
\end{figure}

\ifextend
\sys can deal with transactions initiated by mobile devices in
the same way as described for cross-domain transactions in the previous section.
Once a mobile device sends a request to a remote height$-$1 domain,
the remote domain sends the request
to the local height$-$1 domain and then by using either
the coordinator-based or the optimistic approach,
consensus among both local and remote domains is established.
This, however, requires several rounds of communications across local and remote height$-$1 domains
for every single request initiated by the mobile device in a remote domain.
\fi

In the mobile consensus protocol of \sys,
the local height$-$1 domain shares the state of the mobile node with the remote height$-$1 domain in one round of communication
to enable the remote domain executing transactions initiated by the mobile device.
The state of the node includes the information that is needed to process its
transactions, e.g., the account balance of the node in a micropayment application.

The normal case operation of mobile consensus is presented in algorithm \ref{alg:mobile} and
shown in Figure~\ref{fig:mobilecons} where
$d_l$ and $d_r$ are the local and remote height$-$1 domains.
When the primary $\pi(d_r)$ of the remote domain $d_r$
receives a valid request $m$ from an unauthorized edge device, as shown in lines 5-6,
the primary $\pi(d_r)$ multicasts a signed \hisq message including the request $m$ and its digest $\delta_m$
to nodes of the local domain $d_l$ to obtain the state of the node.
The local domain is the domain where the node is initially registered in.
The primary $\pi(d_r)$ also multicasts the \hisq message to the nodes of its (remote) domain $d_r$ to inform them about request $m$.

Each domain maintains a $lock$ bit for each of its registered edge device to keep track of its mobility.
When an edge device initiates a transaction in a remote domain, the $lock$ is set to {\scriptsize \sf FALSE}, representing that
the state of the edge device in the local domain is outdated.
The domain also defines a variable $remote$ for each edge device to maintain the id of the remote domain that has
the most recent transaction records of the node.
Once the primary $\pi(d_l)$ of the local domain $d_l$ receives a valid \hisq message for its edge device $n$,
as shown in lines 8-9,
it checks the $lock(n)$ to be {\scriptsize \sf TRUE}
(i.e., the state of node $n$ in the local domain is complete and up-to-date) and
then calls {\sc \small GenerateState} function (lines 14-19).
The {\sc \small GenerateState} function constructs the state of mobile node $n$
by executing a predefined application-dependent query on the blockchain.
The primary $\pi(d_l)$ then runs consensus protocol among nodes of the local domain $d_l$ on the state
by sending a message including both \hisq message received from the remote domain $d_r$ as well as
state $H(n)$.
Once consensus is achieved, the primary $\pi(d_l)$ sends a signed \his message including
the extracted state $H(n)$,
the digest $\delta_h$ of the corresponding \hisq message, and
the digest $\delta_m$ of request $m$
to the nodes of the remote domain.
Nodes in $d_l$ also set $lock(n)$ to be {\scriptsize \sf FALSE} and $remote(n)$ to $d_r$.

\newcommand{\hisqc}{{\tiny $\langle\text{\HISQ}, m, \delta_m \rangle_{\sigma_{\pi(d_r)}}$}\xspace}
\newcommand{\hisqcc}{{\tiny $\langle\text{\HISQ}, m, \delta_m \rangle_{\sigma_{\pi(d)}}$}\xspace}
\newcommand{\hisc}{{\tiny $\langle\text{\HIS}, H(n), \delta_h,  \delta_m \rangle_{\sigma}$}\xspace}

\begin{algorithm}[t]
\caption{{\small Mobile Consensus}}
\label{alg:mobile}
\begin{algorithmic}[1]
\State {\em init():} 
\State \quad $i$ := {\em node\_id}
\State \quad $d_l$ := local domain
\State \quad $d_r$ := remote domain
\State upon receiving valid request $m$ from a remote node $n$ and $i$ is $\pi(d_r)$
\State \quad multicast  \hisqc to $d_l$ and $d_r$
\State upon receiving valid \hisqc and $i$ is $\pi(d_l)$
\If{$lock(n)=$ {\tiny \sf TRUE}}
\State \qquad $\pi(d_l)$.\Call{GenerateSate}{$n,d_l,d_r$}
\Else \Comment{{\em $lock(n)$ is {\tiny \sf FALSE} and $remote(n) =d_{r'}$}}
\State \qquad $\pi(d_l)$.\Call{GetSate}{$n,d_l,d_{r'}$}
\State \qquad $\pi(d_l)$.\Call{GenerateSate}{$n,d_l,d_r$}
\EndIf

\Function{GenerateSate}{node $n$, domain $d$, domain $d'$}
\State generate state $H(n)$
\State establish consensus on Sate $H(n)$ among nodes in $d$
\State $lock(n)=$ {\tiny \sf FALSE}, $remote(n) = d'$
\State send \hisc to $d'$
\EndFunction

\Function{GetSate}{node $n$, domain $d$, domain $d'$}
\State send \hisqcc to $d'$
\State $\pi(d')$.\Call{GeneratSate}{$n,d',d$}
\State upon receiving valid \hisc message from $\pi(d')$
\State \quad $lock(n)=$ {\tiny \sf TRUE}
\State \quad establish consensus on transactions of {\scriptsize \sf STATE} message in $d$
\State \quad append the transactions and {\scriptsize \sf commit} message(s) to the ledger
\EndFunction

\end{algorithmic}
\end{algorithm}

If $lock(n)$ is {\scriptsize \sf FALSE} and $remote(n) =d_{r'}$,
some other remote domain $d_{r'}$ has the most recent transaction records.
As a result, as shown in the {\sc \small GetState} function,
the local domain sends a \hisq message to remote domain $d_{r'}$ to obtain
the recent transactions that $n$ has been involved in them.
Upon receiving the state of node $n$ from $d_{r'}$,
the local domain $d_l$, as shown in lines 23-26, establishes consensus on the received state and
updates its blockchain ledger,
Finally, the local domain $d_l$ uses the {\sc \small GenerateState} function
to sends the state of $n$ to the remote domain $d_r$ (line 12).
This situation happens when an edge device
moves to a remote domain $d_{r'}$, initiates transactions
and then moves to another remote domain $d_r$.
In this case, the local domain $d_l$ becomes the intermediary between remote domains $d_r$ and $d_{r'}$ by
obtaining the state from $d_{r'}$, updating its state
and then sending the state to $d_r$.
If the mobile node returns to its local domain,
the local domain updates the ledger and processes the transaction.

\medskip
\noindent {\bf Correctness.}
The correctness of mobile consensus protocol is mainly ensured based on 
the correctness of internal consensus protocols in both local and remote height$-$1 domains.
Assuming the internal consensus protocols are correct, we just need to show that
communications across domains do not violate safety or liveness.
Safety is guaranteed because to send a \his message consensus among
nodes of a domain is needed and
\his messages are certified by the primary of a crash-only domain or $2f+1$ nodes of a Byzantine domain.

To provide liveness, if node $r$ of a domain has not received
a \his message after sending a \hisq message and its timer expires,
the node re-sends the \hisq message to all nodes of the other domain.
The nodes simply re-send the corresponding response if the message has already been processed.
Nodes also log the query messages to detect denial-of-service attacks initiated by malicious nodes.
If the query message is received from a majority of a domain
(they already received the request, line $6$),
the primary will be suspected to be faulty resulting in running the failure handling routine.

If nodes of domain $d'$ receive \hisq from domain $d$,
however, the primary of $d'$ does not initiate consensus on \his message
(after nodes relay the message to the primary),
nodes of $d'$ suspect that the primary is faulty.
Similarly, upon receiving \his messages, 
nodes of domain $d$ wait for the primary of $d$ to initiate consensus.
Otherwise, the primary will be suspected to be faulty.
\section{Experimental Evaluation}\label{sec:exp}

The goal of our evaluations is to measure the impact of
(1) geo-distribution (i.e., nearby domains vs. far apart domains),
(2) cross-domain transactions,
(3) transactions initiated in a remote domain (mobile consensus), and
(4) conflicting transactions (contention in the workload)
in various scenarios on the performance of \sys.

We have implemented a prototype of \sys
and run it on a typical four-level edge network (edge devices, edge servers, fog servers, and cloud servers) structured as a perfect binary tree (following Figure~\ref{fig:edge}). 
Nodes follow either crash or Byzantine failure model.
Each non-leaf domain (except for the last set of experiments) tolerates one failure.
We use Paxos and PBFT as the internal consensus protocol for crash-only and Byzantine domains respectively.

As our experimental workload, we use a micropayment application given that it is a representative and demanding application.
This application uses \sys, 
the blockchain state maintains the balance of each client (edge device),
and clients continuously carry out transactions that
lead to the transfer of financial assets from a sender to a recipient if
all conditions are satisfied, e.g., the sender has a sufficient balance.
The average measured message size (e.g., \req, \one, \pre, \pred, \three, and \reply) is $0.2$ KB while
The \block messages are much larger (depending on the time interval of block propagation and the height of the tree).

The experiments were conducted on
the Amazon EC2 platform on multiple VM instances.
We assigned a separate VM for each node in height$-$1 and above, e.g.,
four VMs are assigned to a domain with Byzantine nodes ($3f+1$).
However, all nodes (clients) of a leaf domain are run on the same VM, i.e.,
we assigned four VMs to the four leaf domains.
Each VM is a c4.2xlarge instance with 8 vCPUs and 15GB RAM,
and an Intel Xeon E5-2666 v3 processor clocked at 3.50 GHz.

\sys follows the edge computing paradigm, 
processes transactions within height$-$1 domains and
propagates \block messages to higher-level domains.
Higher-level domains receive and process \block messages in parallel with
transaction execution within height$-$1 domains.
Given that our goal is to optimize end-user experience at the edge,
we only focus on measuring the end-to-end performance of transaction execution originating with and ending at height$-$1 domains.
It should be noted that since the focus of our evaluation is on  transaction execution in height$-$1 domains,
our evaluation setup does not capture characteristics of edge computing networks, e.g.,
different bandwidth and commuting resources in different network layers.
When reporting throughput measurements, we use an increasing
number of requests until the end-to-end throughput is saturated.

\begin{figure}[t]
\Large
\begin{minipage}{.33\textwidth} \centering
\begin{tikzpicture}[scale=0.49]
\begin{axis}[
    xlabel={Throughput [ktrans/sec]},
    ylabel={Latency [ms]},
    xmin=0, xmax=25,
    ymin=0, ymax=200,
    xtick={0,5,10,15,20},
    ytick={50,100,150},
    legend style={at={(axis cs:0,200)},anchor=north west}, 
    ymajorgrids=true,
    grid style=dashed,
]

\addplot[
    color=green,
    mark=square,
    mark size=4pt,
    line width=0.5mm,
    ]
    coordinates {
  (0.804,7)(3.729,17)(5.613,31)(8.011,45)(12.713,74)(16.911,114)(18.043,147)};

\addplot[
    color=blue,
    mark=+,
    mark size=4pt,
    line width=0.5mm,
    ]
    coordinates {
  (0.888,5)(4.235,17)(8.893,47)(15.642,87)(16.754,95)(17.416,116)};
    
\addplot[
    color=black,
    mark=star,
    mark size=4pt,
    line width=0.5mm,
    ]
    coordinates {
  (0.792,7)(4.235,20)(9.134,44)(14.111,67)(18.154,92)(19.730,115)(20.316,160)};
  
\addplot[
    color=red,
    mark=triangle,
    mark size=4pt,
    line width=0.5mm,
    ]
    coordinates {
  (0.903,5)(4.683,13)(9.302,27)(14.192,45)(18.659,78)(21.711,119)(22.130,140)};
  
\addplot[
    color=magenta,
    mark=otimes,
    mark size=4pt,
    line width=0.5mm,
    ]
    coordinates {
  (0.908,5)(4.704,13)(9.501,26)(14.418,41)(19.203,73)(22.149,109)(22.366,136)};

\addplot[
    color=violet,
    mark=o,
    mark size=4pt,
    line width=0.5mm,
    ]
    coordinates {
  (0.913,5)(4.735,12)(9.834,25)(14.813,39)(19.911,68)(22.654,104)(23.116,134)};

\addlegendentry{AHL}
\addlegendentry{SharPer}
\addlegendentry{Coordinator}
\addlegendentry{Opt-$90\%$ C}
\addlegendentry{Opt-$50\%$ C}
\addlegendentry{Opt-$10\%$ C}
 
\end{axis}
\vspace{-0.5em}
\end{tikzpicture}

{\scriptsize (a) $20\%$ Cross-domain}
\end{minipage}\hfill
\begin{minipage}{.33\textwidth} \centering
\begin{tikzpicture}[scale=0.49]
\begin{axis}[
    xlabel={Throughput [ktrans/sec]},
    ylabel={Latency [ms]},
    xmin=0, xmax=14,
    ymin=0, ymax=600,
    xtick={0,3,6,9,12},
    ytick={150,300,450},
    legend style={at={(axis cs:0,400)},anchor=north west}, 
    ymajorgrids=true,
    grid style=dashed,
]

\addplot[
    color=green,
    mark=square,
    mark size=4pt,
    line width=0.5mm,
    ]
    coordinates {
    (0.065,43)(2.100,82)(3.711,146)(5.330,228)(6.413,292)(7.411,416)};

\addplot[
    color=blue,
    mark=+,
    mark size=4pt,
    line width=0.5mm,
    ]
    coordinates {
   (0.062,48)(2.832,75)(5.087,158)(6.840,234)(8.113,338)(8.613,460)};
    
\addplot[
    color=black,
    mark=star,
    mark size=4pt,
    line width=0.5mm,
    ]
    coordinates {
  (0.063,42)(3.211,70)(5.701,146)(6.735,175)(8.510,241)(9.738,354)(10.121,430)};
  
\addplot[
    color=red,
    mark=triangle,
    mark size=4pt,
    line width=0.5mm,
    ]
    coordinates {
  (0.070,45)(2.409,58)(4.141,74)(5.218,86)(7.049,154)(8.063,232)(8.192,315)};
  
\addplot[
    color=magenta,
    mark=otimes,
    mark size=4pt,
    line width=0.5mm,
    ]
    coordinates {
  (0.070,33)(3.281,50)(5.561,75)(7.109,100)(9.619,160)(10.241,242)(10.262,281)};

\addplot[
    color=violet,
    mark=o,
    mark size=4pt,
    line width=0.5mm,
    ]
    coordinates {
  (0.070,21)(4.023,44)(7.034,73)(9.213,112)(10.920,151)(12.830,266)(13.202,336)};
  
% \addlegendentry{AHL}
% \addlegendentry{SharPer}
% \addlegendentry{Coordinator}
% \addlegendentry{Opt-$90\%$ C}
% \addlegendentry{Opt-$50\%$ C}
% \addlegendentry{Opt-$10\%$ C}
 
\end{axis}
\vspace{-0.5em}
\end{tikzpicture}

{\scriptsize (b) $80\%$ Cross-domain}
\end{minipage}\hfill
\begin{minipage}{.33\textwidth} \centering
\begin{tikzpicture}[scale=0.49]
\begin{axis}[
    xlabel={Throughput [ktrans/sec]},
    ylabel={Latency [ms]},
    xmin=0, xmax=11.5,
    ymin=0, ymax=800,
    xtick={0,2,4,6,8,10},
    ytick={200,400,600},
    legend style={at={(axis cs:0,400)},anchor=north west}, 
    ymajorgrids=true,
    grid style=dashed,
]

\addplot[
    color=green,
    mark=square,
    mark size=4pt,
    line width=0.5mm,
    ]
    coordinates {
   (0.067,56)(1.404,74)(3.012,183)(4.110,293)(4.854,367)(5.407,453)(6.311,670)};

\addplot[
    color=blue,
    mark=+,
    mark size=4pt,
    line width=0.5mm,
    ]
    coordinates {
    (0.058,58)(1.601,73)(2.703,120)(4.320,222)(5.431,325)(6.772,447)(7.412,612)};
    
\addplot[
    color=black,
    mark=star,
    mark size=4pt,
    line width=0.5mm,
    ]
    coordinates {
    (0.064,60)(1.765,74)(3.120,103)(5.318,201)(7.532,321)(8.772,429)(9.102,567)};
  
\addplot[
    color=red,
    mark=triangle,
    mark size=4pt,
    line width=0.5mm,
    ]
    coordinates {
    (0.060,55)(1.348,80)(2.405,130)(3.177,173)(3.899,270)(4.115,403)};
    
\addplot[
    color=magenta,
    mark=otimes,
    mark size=4pt,
    line width=0.5mm,
    ]
    coordinates {
    (0.063,40)(2.502,67)(4.191,114)(5.602,169)(8.442,284)(8.692,351)};
    
\addplot[
    color=violet,
    mark=o,
    mark size=4pt,
    line width=0.5mm,
    ]
    coordinates {
    (0.066,19)(3.720,54)(6.411,109)(8.042,142)(10.011,240)(10.854,390)};

% \addlegendentry{AHL}
% \addlegendentry{SharPer}
% \addlegendentry{Coordinator}
% \addlegendentry{Opt-$90\%$ C}
% \addlegendentry{Opt-$50\%$ C}
% \addlegendentry{Opt-$10\%$ C}

\end{axis}
\vspace{-0.5em}
\end{tikzpicture}

{\scriptsize (c) $100\%$ Cross-domain}
\end{minipage}
\caption{Cross-Domain Transactions (Crash-only)}
  \label{fig:cross-c}
\end{figure}
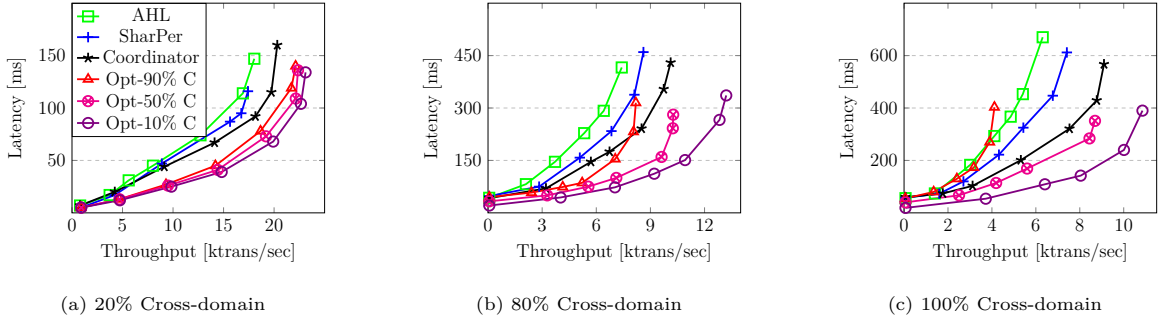

\subsection{Cross-Domain Transactions}
In the first set of experiments, we evaluate \sys in
workloads with different percentages of cross-domain transactions (i.e., $0\%$, $20\%$, $80\%$, and $100\%$).
Domains are distributed over four nearby AWS regions,
i.e., Frankfurt ({\em FR}), Milan ({\em MI}), London ({\em LDN}), and Paris ({\em PAR}) where
the average measured Round-Trip Time (RTT) between every pair of Amazon data centers is as follows;
{\em FR} $\rightleftharpoons$ {\em MI}: $11$ ms,
{\em FR} $\rightleftharpoons$ {\em LDN}: $17$ ms,
{\em FR} $\rightleftharpoons$ {\em PAR}: $9$ ms,
{\em MI} $\rightleftharpoons$ {\em LDN}: $25$ ms,
{\em MI} $\rightleftharpoons$ {\em PAR}: $19$ ms, and
{\em LDN} $\rightleftharpoons$ {\em PAR}: $10$ ms.
In this scenario, each leaf and its corresponding height$-$1 domain is placed in
one of the $4$ data centers, while the higher-level domains are in the {\em FR} region.

We compare the coordinator-based and optimistic protocols of \sys with
scalable solutions SharPer \cite{amiri2021sharper}, and AHL \cite{dang2018towards}.
SharPer and AHL are chosen because the experiments focus on studying the impact of
hierarchical structure on processing cross-domain transactions.
Due to the emphasis of the experiments, we only implemented the cross-shard consensus protocol of AHL
where a reference committee uses 2PC to order transactions (without using trusted hardware).
The internal transactions of all approaches are processed in the same way using Paxos (for crash-only domains) or PBFT (for Byzantine domains) protocol.
Both SharPer and AHL are run over a network with four clusters (domains) with $f=1$ (the same setting as height$-$1 of \sys).
For a fair comparison, the latency in \sys is measured from the initiation of a transaction to when it gets committed to the blockchain of height$-$1 domain(s).
Two randomly chosen domains (sender and recipient) are involved in each transaction. 

\begin{figure}[t!]
\large
\begin{minipage}{.33\textwidth} \centering
\begin{tikzpicture}[scale=0.49]
\begin{axis}[
    xlabel={Throughput [ktrans/sec]},
    ylabel={Latency [ms]},
    xmin=0, xmax=20,
    ymin=0, ymax=400,
    xtick={0,4,8,12,16},
    ytick={100,200,300},
    legend style={at={(axis cs:0,400)},anchor=north west}, 
    ymajorgrids=true,
    grid style=dashed,
]

\addplot[
    color=green,
    mark=square,
    mark size=4pt,
    line width=0.5mm,
    ]
    coordinates {
    (0.801,10)(2.345,20)(5.011,41)(10.123,108)(13.492,195)(13.542,243)};

\addplot[
    color=blue,
    mark=+,
    mark size=4pt,
    line width=0.5mm,
    ]
    coordinates {
    (0.836,7)(3.781,19)(7.841,47)(11.121,94)(14.409,174)(16.211,279)};
    
\addplot[
    color=black,
    mark=star,
    mark size=4pt,
    line width=0.5mm,
    ]
    coordinates {
    (0.906,7)(4.681,15)(8.613,38)(12.711,80)(16.042,117)(17.013,224)};

\addplot[
    color=red,
    mark=triangle,
    mark size=4pt,
    line width=0.5mm,
    ]
    coordinates {
    (0.901,7)(5.605,9)(9.402,30)(13.403,68)(16.792,95)(17.392,160)};

\addplot[
    color=magenta,
    mark=otimes,
    mark size=4pt,
    line width=0.5mm,
    ]
    coordinates {
    (0.911,7)(5.730,9)(9.650,29)(13.752,60)(17.101,91)(18.354,203)};
    
\addplot[
    color=violet,
    mark=o,
    mark size=4pt,
    line width=0.5mm,
    ]
    coordinates {
    (0.915,7)(5.802,9)(9.821,27)(14.003,55)(17.503,89)(18.232,140)};

\addlegendentry{AHL}
\addlegendentry{SharPer}
\addlegendentry{Coordinator}
\addlegendentry{Opt-$90\%$ C}
\addlegendentry{Opt-$50\%$ C}
\addlegendentry{Opt-$10\%$ C}

\end{axis}
\vspace{-0.5em}
\end{tikzpicture}

{\scriptsize (a) $20\%$ Cross-domain}
\end{minipage}\hfill
\begin{minipage}{.33\textwidth} \centering
\begin{tikzpicture}[scale=0.49]
\begin{axis}[
    xlabel={Throughput [ktrans/sec]},
    ylabel={Latency [ms]},
    xmin=0, xmax=12,
    ymin=0, ymax=1200,
    xtick={0,2,4,6,8,10},
    ytick={300,600,900},
    legend style={at={(axis cs:0,1200)},anchor=north west}, 
    ymajorgrids=true,
    grid style=dashed,
]

\addplot[
    color=green,
    mark=square,
    mark size=4pt,
    line width=0.5mm,
    ]
    coordinates {
    (0.51,88)(1.43,171)(2.201,260)(4.724,593)(5.432,889)};

\addplot[
    color=blue,
    mark=+,
    mark size=4pt,
    line width=0.5mm,
    ]
    coordinates {
    (0.49,59)(0.912,85)(2.943,209)(4.523,419)(5.673,705)(6.032,919)};
    
\addplot[
    color=black,
    mark=star,
    mark size=4pt,
    line width=0.5mm,
    ]
    coordinates {
    (0.50,49)(1.131,70)(3.001,175)(5.131,347)(6.791,550)(8.151,957)(8.231,1093)};
  
\addplot[
    color=red,
    mark=triangle,
    mark size=4pt,
    line width=0.5mm,
    ]
    coordinates {
    (0.43,52)(0.632,65)(1.803,92)(2.428,124)(3.243,165)(3.866,304)(4.702,535)};
    
\addplot[
    color=magenta,
    mark=otimes,
    mark size=4pt,
    line width=0.5mm,
    ]
    coordinates {
    (0.45,39)(0.975,63)(2.473,88)(4.102,123)(5.201,163)(6.103,302)(7.332,515)(7.611,653)};
    
\addplot[
    color=violet,
    mark=o,
    mark size=4pt,
    line width=0.5mm,
    ]
    coordinates {
    (0.49,34)(1.221,47)(3.321,88)(5.882,120)(7.231,153)(8.730,249)(10.601,392)(11.281,468)(11.410,687)};

% \addlegendentry{AHL}
% \addlegendentry{SharPer}
% \addlegendentry{Coordinator}
% \addlegendentry{Opt-$90\%$ C}
% \addlegendentry{Opt-$50\%$ C}
% \addlegendentry{Opt-$10\%$ C}

\end{axis}
\vspace{-0.5em}
\end{tikzpicture}

{\scriptsize (b) $80\%$ Cross-domain}
\end{minipage}\hfill
\begin{minipage}{.33\textwidth} \centering
\begin{tikzpicture}[scale=0.49]
\begin{axis}[
    xlabel={Throughput [ktrans/sec]},
    ylabel={Latency [ms]},
    xmin=0, xmax=10,
    ymin=0, ymax=1500,
    xtick={0,2,4,6,8},
    ytick={400,800,1200},
    legend style={at={(axis cs:0,1000)},anchor=north west}, 
    ymajorgrids=true,
    grid style=dashed,
]

\addplot[
    color=green,
    mark=square,
    mark size=4pt,
    line width=0.5mm,
    ]
    coordinates {
    (0.44,116)(1.02,202)(1.841,371)(3.719,692)(4.571,973)(4.683,1342)};

\addplot[
    color=blue,
    mark=+,
    mark size=4pt,
    line width=0.5mm,
    ]
    coordinates {
    (0.50,54)(0.954,102)(1.394,175)(2.402,404)(4.242,853)(5.242,1432)};
    
\addplot[
    color=black,
    mark=star,
    mark size=4pt,
    line width=0.5mm,
    ]
    coordinates {
    (0.51,43)(0.962,73)(2.027,157)(3.114,231)(4.701,502)(5.711,784)(6.331,1203)};

\addplot[
    color=red,
    mark=triangle,
    mark size=4pt,
    line width=0.5mm,
    ]
    coordinates {
    (0.423,70)(0.905,90)(1.665,140)(2.155,210)(2.749,350)(3.451,510)(3.586,830)};

\addplot[
    color=magenta,
    mark=otimes,
    mark size=4pt,
    line width=0.5mm,
    ]
    coordinates {
    (0.529,17)(1.055,91)(2.192,119)(3.232,170)(4.103,257)(5.645,497)(5.813,762)};
    
\addplot[
    color=violet,
    mark=o,
    mark size=4pt,
    line width=0.5mm,
    ]
    coordinates {
    (0.59,70)(1.221,82)(2.844,93)(4.351,121)(5.815,152)(7.634,290)(9.540,540)(9.713,801)};

% \addlegendentry{AHL}
% \addlegendentry{SharPer}
% \addlegendentry{Coordinator}
% \addlegendentry{Opt-$90\%$ C}
% \addlegendentry{Opt-$50\%$ C}
% \addlegendentry{Opt-$10\%$ C}

\end{axis}
\vspace{-0.5em}
\end{tikzpicture}

{\scriptsize (c) $100\%$ Cross-domain}
\end{minipage}
\caption{Cross-Domain Transactions (Byzantine)}
\label{fig:cross-b}
\end{figure}
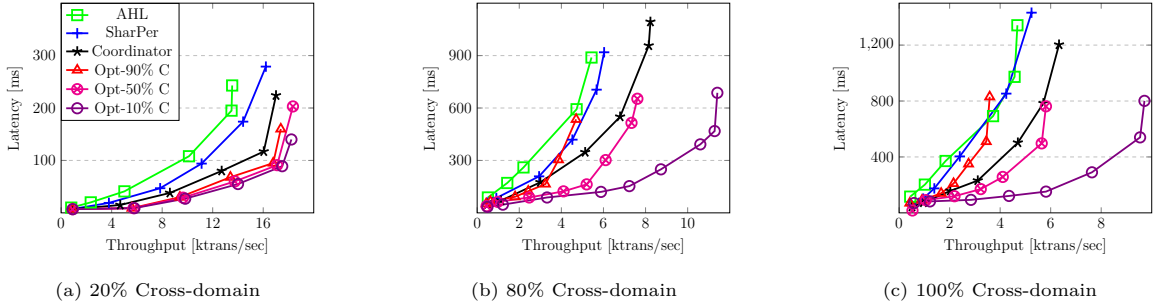

In the optimistic approach, as discussed in Sec~\ref{sec:opt}, a cross-domain transaction might be aborted due to inconsistency, i.e.,
two concurrent cross-domain transactions have been appended to the ledger of two domains in a different order,
resulting in aborting all their data-dependent transactions.
To measure the effects of contention on the performance of the optimistic protocol,
we consider three workloads with different degrees of contention between transactions of each domain, i.e.,
$10\%$ (the default value for all workloads), $50\%$, and $90\%$ read-write conflicts,
Figures~\ref{fig:cross-c} and \ref{fig:cross-b} demonstrate the results with crash-only and Byzantine domains.

When all nodes are crash-only and all transactions are internal, 
\sys is able to process more than $31000$ tps with $100$ ms latency.
In this scenario,
each domain processes its transactions independently and
the throughput of the entire system will increase linearly with the number of domains.
With $20\%$ cross-domain transactions, as shown in Figure~\ref{fig:cross-c}(a), the optimistic approach with $10\%$ contention 
shows the best performance by processing $22500$ tps with $105$ ms latency.
This is expected because the optimistic approach does not require any communication
across domains.
In this scenario, only $0.17\%$ of transactions were appended to the ledgers in an inconsistent order, hence,
increasing the percentage of contention in the workload to $50\%$ and $90\%$ (Opt-$50\%$ C and Opt-$90\%$ C graphs)
does not significantly affect the performance of the optimistic protocol.
The coordinator-based approach also processes $19700$ tps with $115$ ms latency which is $17\%$ more
than AHL ($16900$ tps with the same latency).

Increasing the percentage of cross-shard transactions to $80\%$ and $100\%$, as shown in Figure~\ref{fig:cross-c}(b) and (c),
results in a larger performance gap between the coordinator-based approach and the existing systems (SharPer and AHL), e.g.,
in the workload with $100\%$ cross-domain transaction,
the coordinator-based approach processes $63\%$ transactions more than the AHL with the same latency.
This is expected because in AHL, the single coordinator becomes overloaded by cross-domain transactions and
in SharPer, consensus across domains becomes a bottleneck.
However, \sys processes transactions efficiently by relying on multiple coordinator domains.
The optimistic approach demonstrates lower performance in workloads with $50\%$ and $90\%$ contention
due to higher inconsistencies.

In the presence of Byzantine nodes, as shown in Fig~\ref{fig:cross-b}, \sys shows similar behavior,
although with lower throughput and higher latency (due to the higher cost of BFT protocols compared to CFT protocols).

\subsection{Transactions Initiated by Mobile Devices}

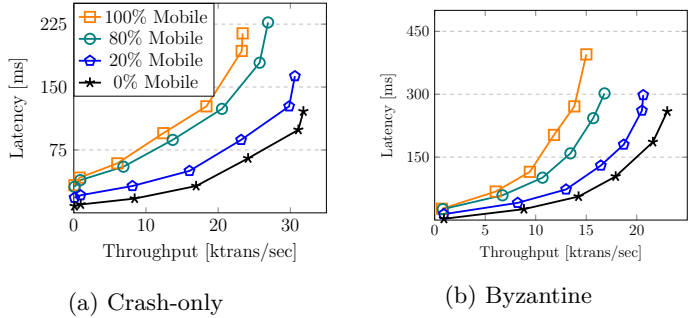
\begin{figure}[t]
\Large
\centering
\begin{minipage}{.23\textwidth}\centering
\begin{tikzpicture}[scale=0.49]
\begin{axis}[
    xlabel={Throughput [ktrans/sec]},
    ylabel={Latency [ms]},
    xmin=0, xmax=35,
    ymin=0, ymax=250,
    xtick={0,10,20,30},
    ytick={75,150,225},
    legend style={at={(axis cs:0,250)},anchor=north west}, 
    ymajorgrids=true,
    grid style=dashed,
]

\addplot[
    color=orange,
    mark=square,
    mark size=4pt,
    line width=0.5mm,
    ]
    coordinates {
    (0.093,33)(0.813,42)(6.031,59)(12.404,95)(18.376,127)(23.265,193)(23.392,214)};
    
\addplot[
    color=teal,
    mark=o,
    mark size=4pt,
    line width=0.5mm,
    ]
    coordinates {
    (0.095,31)(0.901,39)(6.870,55)(13.701,87)(20.529,124)(25.748,179)(26.913,227)};

 \addplot[
    color=blue,
    mark=pentagon,
    mark size=4pt,
    line width=0.5mm,
    ]
    coordinates {
    (0.097,18)(0.934,21)(8.119,32)(16.021,50)(23.182,87)(29.834,127)(30.635,163)};

\addplot[
    color=black,
    mark=star,
    mark size=4pt,
    line width=0.5mm,
    ]
    coordinates {
    (0.099,8)(0.958,10)(8.432,17)(16.950,32)(24.150,65)(31.114,99)(31.810,121)};

\addlegendentry{$100\%$ Mobile}
\addlegendentry{$80\%$ Mobile}
\addlegendentry{$20\%$ Mobile}
\addlegendentry{$0\%$ Mobile}
 
\end{axis}
\end{tikzpicture}

{\footnotesize (a) Crash-only}
\end{minipage}\hspace{2em}
\begin{minipage}{.23\textwidth} \centering
\large
\centering
\begin{tikzpicture}[scale=0.49]
\begin{axis}[
    xlabel={Throughput [ktrans/sec]},
    ylabel={Latency [ms]},
    xmin=0, xmax=25,
    ymin=0, ymax=500,
    xtick={0,5,10,15,20},
    ytick={150,300,450},
    legend style={at={(axis cs:0,500)},anchor=north west}, 
    ymajorgrids=true,
    grid style=dashed,
]

\addplot[
    color=orange,
    mark=square,
    mark size=4pt,
    line width=0.5mm,
    ]
    coordinates {
    (0.719,27)(6.042,68)(9.423,115)(11.811,203)(13.819,271)(15.011,395)};
    
\addplot[
    color=teal,
    mark=o,
    mark size=4pt,
    line width=0.5mm,
    ]
    coordinates {
    (0.804,26)(6.711,59)(10.6722,101)(13.435,159)(15.691,243)(16.813,302)};

 \addplot[
    color=blue,
    mark=pentagon,
    mark size=4pt,
    line width=0.5mm,
    ]
    coordinates {
    (0.892,14)(8.211,41)(13.002,73)(16.443,130)(18.711,180)(20.513,261)(20.642,298)};

\addplot[
    color=black,
    mark=star,
    mark size=4pt,
    line width=0.5mm,
    ]
    coordinates {    
    (0.913,3)(8.832,26)(14.241,56)(17.913,104)(21.601,186)(23.017,259)};

% \addlegendentry{$100\%$ Mobile}
% \addlegendentry{$80\%$ Mobile}
% \addlegendentry{$20\%$ Mobile}
% \addlegendentry{$0\%$ Mobile}

\end{axis}
\end{tikzpicture}

{\footnotesize (b) Byzantine}
\end{minipage}
\caption{Performance with Mobile Devices}
\label{fig:mobile}
\end{figure}

In the second set of experiments, we measure the performance of 
the mobile consensus protocol to process remotely initiated transactions.
The network is the same as Section~\ref{sec:exp} and
we consider four workloads with different percentages (i.e., $0\%$, $20\%$, $80\%$, and $100\%$)
of mobile nodes where
a local and a remote height$-$1 domains are involved in each mobile transaction.
To simulate the mobility of edge devices,
we run an instance of each edge device within the VM of all leaf domains (data centers).
A mobile node initiates $10$ transactions within the remote domain before moving back to its local domain.
The state in a micropayment application includes the balance of the mobile node.
Figure~\ref{fig:mobile}(a) and Figure~\ref{fig:mobile}(b) show the results with crash-ony and Byzantine domains.

With crash-only nodes and local transactions, \sys, as shown in Figure~\ref{fig:mobile}(a),
processes $31000$ tps with less than $100$ ms latency
(same as $0\%$ cross-domain transaction).
Adding $20\%$ mobile transactions, \sys still processes $29800$ transactions (only $\sim 4\%$ reduction).
Similarly, with $80\%$ and $100\%$ mobile transactions,
\sys processes $25700$ and $23200$ tps.
This demonstrates the effectiveness of \sys in handling mobile devices:
increasing the percentage of mobile devices from $0\%$ to $100\%$
results in only a $25\%$ reduction in throughput.
\sys demonstrates similar behavior with Byzantine domains (Figure~\ref{fig:mobile}(b)).
However, since establishing consensus on \his messages is more expensive with Byzantine nodes,
\sys incurs $36\%$ reduction in throughput by increasing the percentage of mobile devices from $0\%$ to $100\%$.
These results clearly demonstrate the capability of \sys in supporting applications that requires mobility of nodes, e.g., ridesharing.

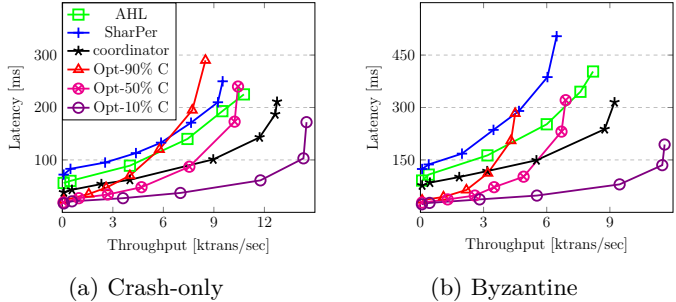
\begin{figure}[h]
\large
\centering
\begin{minipage}{.23\textwidth} \centering
\begin{tikzpicture}[scale=0.49]
\begin{axis}[
    xlabel={Throughput [ktrans/sec]},
    ylabel={Latency [ms]},
    xmin=0, xmax=15,
    ymin=0, ymax=400,
    xtick={0,3,6,9,12},
    ytick={100,200,300},
    legend style={at={(axis cs:0,400)},anchor=north west}, 
    ymajorgrids=true,
    grid style=dashed,
]

\addplot[
    color=green,
    mark=square,
    mark size=4pt,
    line width=0.5mm,
    ]
    coordinates {
    (0.08,56)(0.485,60)(4.012,89)(7.431,140)(9.502,193)(10.784,225)};
    
\addplot[
    color=blue,
    mark=+,
    mark size=4pt,
    line width=0.5mm,
    ]
    coordinates {
    (0.08,72)(0.485,83)(2.542,95)(4.372,113)(5.863,133)(7.644,171)(9.239,210)(9.521,250)};
    
\addplot[
    color=black,
    mark=star,
    mark size=4pt,
    line width=0.5mm,
    ]
    coordinates {
    (0.09,38)(0.573,43)(2.321,54)(4.012,62)(8.963,101)(11.716,143)(12.643,187)(12.751,211)};

\addplot[
    color=red,
    mark=triangle,
    mark size=4pt,
    line width=0.5mm,
    ]
    coordinates {(0.09,23)(1.573,35)(2.603,47)(4.012,70)(5.765,120)(7.719,195)(8.501,290)};
  
\addplot[
    color=magenta,
    mark=otimes,
    mark size=4pt,
    line width=0.5mm,
    ]
    coordinates {(0.07,19)(0.980,27)(2.703,34)(4.702,48)(7.544,87)(10.231,173)(10.433,240)};

\addplot[
    color=violet,
    mark=o,
    mark size=4pt,
    line width=0.5mm,
    ]
    coordinates {(0.09,17)(0.573,21)(3.603,27)(7.012,37)(11.765,61)(14.319,103)(14.501,172)};

\addlegendentry{AHL}
\addlegendentry{SharPer}
\addlegendentry{coordinator}
\addlegendentry{Opt-$90\%$ C}
\addlegendentry{Opt-$50\%$ C}
\addlegendentry{Opt-$10\%$ C}

\end{axis}
\end{tikzpicture}

{\footnotesize (a) Crash-only}
\end{minipage}\hspace{2em}
\begin{minipage}{.23\textwidth} \centering
\begin{tikzpicture}[scale=0.49]
\begin{axis}[
    xlabel={Throughput [ktrans/sec]},
    ylabel={Latency [ms]},
    xmin=0, xmax=12,
    ymin=0, ymax=600,
    xtick={0,3,6,9},
    ytick={150,300,450},
    legend style={at={(axis cs:0,600)},anchor=north west}, 
    ymajorgrids=true,
    grid style=dashed,
]

\addplot[
    color=green,
    mark=square,
    mark size=4pt,
    line width=0.5mm,
    ]
    coordinates {
    (0.08,91)(0.394,108)(3.184,163)(6.001,252)(7.601,345)(8.201,403)};

\addplot[
    color=blue,
    mark=+,
    mark size=4pt,
    line width=0.5mm,
    ]
    coordinates {
    (0.08,124)(0.394,137)(1.974,168)(3.471,236)(4.682,290)(6.034,387)(6.471,504)};
    
\addplot[
    color=black,
    mark=star,
    mark size=4pt,
    line width=0.5mm,
    ]
    coordinates {
    (0.08,77)(0.452,85)(1.845,101)(3.173,118)(5.511,149)(8.754,238)(9.221,315)};

\addplot[
    color=red,
    mark=triangle,
    mark size=4pt,
    line width=0.5mm,
    ]
    coordinates {
    (0.08,35)(1.094,44)(2.184,64)(3.201,112)(4.301,205)(4.501,283)};
  
\addplot[
    color=magenta,
    mark=otimes,
    mark size=4pt,
    line width=0.5mm,
    ]
    coordinates {
    (0.08,28)(1.294,37)(2.584,48)(3.501,72)(4.901,102)(6.701,231)(6.901,321)};

\addplot[
    color=violet,
    mark=o,
    mark size=4pt,
    line width=0.5mm,
    ]
    coordinates {
    (0.07,23)(0.433,27)(2.811,37)(5.532,48)(9.462,80)(11.489,135)(11.600,194)};

% \addlegendentry{AHL}
% \addlegendentry{SharPer}
% \addlegendentry{coordinator}
% \addlegendentry{Opt-$90\%$ C}
% \addlegendentry{Opt-$50\%$ C}
% \addlegendentry{Opt-$10\%$ C}

\end{axis}
\end{tikzpicture}

{\footnotesize (b) Byzantine}
\end{minipage}
\caption{Wide Area ($10\%$ cross-domain)}
  \label{fig:distance}
\end{figure}

\subsection{Scalability Over Wide-Area Domains}\label{sec:widearea}

In the next experiments, the impact of long network distance on the performance of \sys is measured.
We distribute domains over $7$ far apart AWS regions all around the world,
i.e., California ({\em CA}), Oregon ({\em OR}), Virginia ({\em VA}), Ohio ({\em OH}),
Tokyo ({\em TY}), Seoul ({\em SU}), and Hong Kong ({\em HK})\footnote{The average measured Round-Trip Time (RTT)
between every pair of Amazon data centers can be found at https://www.cloudping.co/grid}.
In this scenario, each leaf and its corresponding height$-$1 domain is placed in
one of the {\em TY}, {\em HK}, {\em VA}, and {\em OH} data-centers,
the height$-$2 domains are in {\em SU} and {\em OR} and the root domain is in the {\em CA} region.
Nodes of the same domain are placed in a single AWS region to simulate the behavior of edge networks, i.e.,
edge devices (servers) are within a small geographical domain.
We consider workloads with $90\%$ internal and $10\%$ cross-domain transactions
(typical settings in partitioned datastores \cite{taft2014store}) where
two randomly chosen domains are involved in each cross-domain transaction.
Figures~\ref{fig:distance}(a) and \ref{fig:distance}(b)
depict the results for crash-only and Byzantine domains.

As shown in Figure~\ref{fig:distance}(a),
the optimistic protocol in the low contention workload still has the best performance
(note that the workload includes only $10\%$ cross-domain transactions).
However, conflicting transactions significantly reduce the performance of the optimistic protocol
in high contention workloads ({\small \sf Opt-$50\%$C} and {\small \sf Opt-$90\%$C}) compared to nearby domains (Figure~\ref{fig:cross-c}).
This is expected because when domains are far apart, resolving inconsistencies requires more time
resulting in aborting more data-dependent transactions.
Furthermore, the gap between the performance of the coordinator-based approach and
AHL (single coordinator) has been increased, demonstrating the
effectiveness of the coordinator-based approach over wide-area networks.
Interestingly, AHL demonstrates better performance compared to SharPer
because SharPer requires rounds of communication among nodes of domains over a wide area.
In the presence of Byzantine domains, as shown in Figure~\ref{fig:distance}(b),
all protocols demonstrate similar behavior as the previous case.

\begin{figure}[h]
\large
\centering
\begin{minipage}{.23\textwidth}\centering
\begin{tikzpicture}[scale=0.49]
\begin{axis}[
    xlabel={Throughput [ktrans/sec]},
    ylabel={Latency [ms]},
    xmin=0, xmax=35,
    ymin=0, ymax=600,
    xtick={0,10,20,30},
    ytick={150,300,450},
    legend style={at={(axis cs:0,600)},anchor=north west}, 
    ymajorgrids=true,
    grid style=dashed,
]

\addplot[
    color=orange,
    mark=square,
    mark size=4pt,
    line width=0.5mm,
    ]
    coordinates {
    (0.661,171)(4.983,190)(10.292,211)(13.101,240)(16.589,307)(19.011,493)(19.212,544)};
    
\addplot[
    color=teal,
    mark=o,
    mark size=4pt,
    line width=0.5mm,
    ]
    coordinates {
    (0.082,130)(0.774,133)(5.776,154)(11.05,186)(17.426,267)(21.875,362)(22.011,510)};

 \addplot[
    color=blue,
    mark=pentagon,
    mark size=4pt,
    line width=0.5mm,
    ]
    coordinates {
    (0.089,37)(0.744,51)(7.047,78)(13.932,108)(20.166,147)(25.932,211)(26.102,290)};

\addplot[
    color=black,
    mark=star,
    mark size=4pt,
    line width=0.5mm,
    ]
    coordinates {
    (0.099,8)(0.958,10)(8.432,17)(16.950,32)(24.150,65)(31.114,99)(31.810,121)};

\addlegendentry{$100\%$ Mobile}
\addlegendentry{$80\%$ Mobile}
\addlegendentry{$20\%$ Mobile}
\addlegendentry{$0\%$ Mobile}
 
\end{axis}
\end{tikzpicture}

{\footnotesize (a) Crash-only}
\end{minipage}\hspace{2em}
\begin{minipage}{.23\textwidth} \centering
\begin{tikzpicture}[scale=0.49]
\begin{axis}[
    xlabel={Throughput [ktrans/sec]},
    ylabel={Latency [ms]},
    xmin=0, xmax=25,
    ymin=0, ymax=700,
    xtick={0,5,10,15,20},
    ytick={200,400,600},
    legend pos=north west,
    ymajorgrids=true,
    grid style=dashed,
]

\addplot[
    color=orange,
    mark=square,
    mark size=4pt,
    line width=0.5mm,
    ]
    coordinates {
    (0.545,194)(5.301,240)(8.432,313)(10.421,384)(11.502,456)(12.019,531)(12.391,627)};
    
\addplot[
    color=teal,
    mark=o,
    mark size=4pt,
    line width=0.5mm,
    ]
    coordinates {
    (0.819,140)(2.561,160)(5.819,190)(7.991,230)(10.281,300)(11.619,341)(13.222,460)(13.615,562)};

 \addplot[
    color=blue,
    mark=pentagon,
    mark size=4pt,
    line width=0.5mm,
    ]
    coordinates {
    (0.819,40)(3.256,58)(6.354,86)(9.751,130)(10.995,161)(13.401,200)(17.343,310)(17.731,390)};

\addplot[
    color=black,
    mark=star,
    mark size=4pt,
    line width=0.5mm,
    ]
    coordinates {
    (0.913,3)(8.832,26)(14.241,56)(17.913,104)(22.601,216)(23.017,289)};
 
\end{axis}
\end{tikzpicture}

{\footnotesize (b) Byzantine}
\end{minipage}
\caption{Performance with Mobile Devices over Wide Area}
\label{fig:distance-mob}
\end{figure}
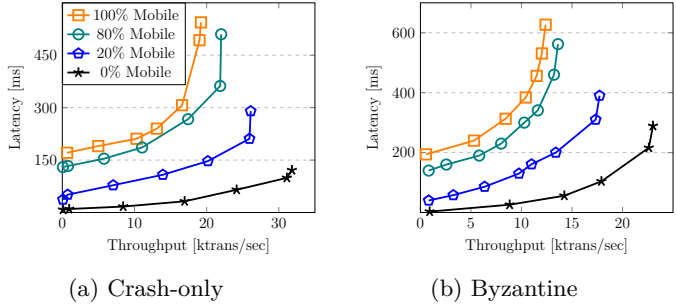

We then use the same settings to measure the impact of network distance on mobile transactions in
workloads with $0\%$, $20\%$, $80\%$, and $100\%$ mobile nodes.
As before, each leaf, i.e., edge devices, and its corresponding height$-$1 domain is placed in
one of the {\em TY}, {\em HK}, {\em VA}, and {\em OH} data-centers.
As shown in Figure~\ref{fig:distance-mob}(a), while processing mobile transactions over a wide area results in higher latency,
\sys still demonstrate an efficient throughput:
when the percentage of mobile devices increases from $0\%$ to $100\%$,
\sys incurs only a $38\%$ reduction in its throughput (with crash-only nodes).

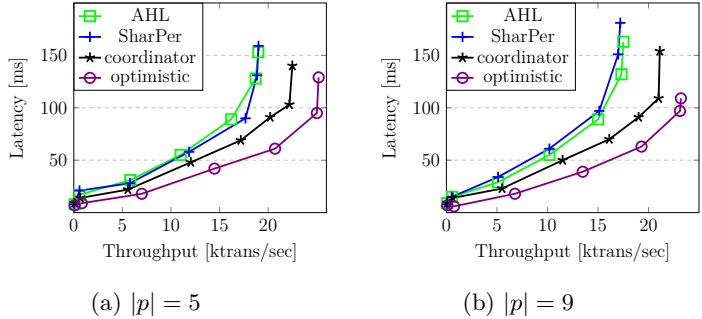
\begin{figure}[h]
\Large
\centering
\begin{minipage}{.23\textwidth} \centering
\begin{tikzpicture}[scale=0.49]
\begin{axis}[
    xlabel={Throughput [ktrans/sec]},
    ylabel={Latency [ms]},
    xmin=0, xmax=26,
    ymin=0, ymax=200,
    xtick={0,5,10,15,20},
    ytick={50,100,150},
    legend style={at={(axis cs:0,200)},anchor=north west}, 
    ymajorgrids=true,
    grid style=dashed,
]

\addplot[
    color=green,
    mark=square,
    mark size=4pt,
    line width=0.5mm,
    ]
    coordinates {
    (0.08,8)(0.588,17)(5.819,31)(11.001,55)(16.203,89)(18.714,128)(18.989,153)};

\addplot[
    color=blue,
    mark=+,
    mark size=4pt,
    line width=0.5mm,
    ]
    coordinates {
    (0.08,8)(0.588,21)(5.769,28)(11.871,58)(17.673,90)(18.862,131)(19.010,159)};
  
\addplot[
    color=black,
    mark=star,
    mark size=4pt,
    line width=0.5mm,
    ]
    coordinates {
    (0.08,9)(0.664,14)(5.572,22)(12.041,48)(17.191,69)(20.211,91)(22.194,103)(22.493,140)};

\addplot[
    color=violet,
    mark=o,
    mark size=4pt,
    line width=0.5mm,
    ]
    coordinates {
    (0.09,7)(0.832,9)(7.001,18)(14.481,42)(20.712,61)(25.019,95)(25.200,129)};

\addlegendentry{AHL}
\addlegendentry{SharPer}
\addlegendentry{coordinator}
\addlegendentry{optimistic}
 
\end{axis}
\end{tikzpicture}

{\footnotesize (a) $|p|=5$}
\end{minipage}\hspace{2em}
\begin{minipage}{.23\textwidth} \centering
\begin{tikzpicture}[scale=0.49]
\begin{axis}[
    xlabel={Throughput [ktrans/sec]},
    ylabel={Latency [ms]},
    xmin=0, xmax=25,
    ymin=0, ymax=200,
    xtick={0,5,10,15,20},
    ytick={50,100,150},
    legend style={at={(axis cs:0,200)},anchor=north west}, 
    ymajorgrids=true,
    grid style=dashed,
]

\addplot[
    color=green,
    mark=square,
    mark size=4pt,
    line width=0.5mm,
    ]
    coordinates {
    (0.08,9)(0.559,15)(5.104,29)(10.201,55)(15.011,89)(17.324,132)(17.509,163)};

\addplot[
    color=blue,
    mark=+,
    mark size=4pt,
    line width=0.5mm,
    ]
    coordinates {
    (0.08,8)(0.519,15)(5.119,34)(10.191,61)(15.131,97)(17.001,151)(17.210,181)};
  
\addplot[
    color=black,
    mark=star,
    mark size=4pt,
    line width=0.5mm,
    ]
    coordinates {
    (0.08,9)(0.622,14)(5.511,23)(11.503,50)(16.119,70)(19.019,91)(21.001,109)(21.119,154)};

\addplot[
    color=violet,
    mark=o,
    mark size=4pt,
    line width=0.5mm,
    ]
    coordinates {
    (0.09,7)(0.753,6)(6.783,18)(13.492,39)(19.311,63)(23.119,97)(23.210,109)};

\addlegendentry{AHL}
\addlegendentry{SharPer}
\addlegendentry{coordinator}
\addlegendentry{optimistic}

\end{axis}
\end{tikzpicture}

{\footnotesize (b) $|p|=9$}
\end{minipage}
\caption{Increasing the Number of Crash-Only Nodes}
  \label{fig:nodes-c}
\end{figure}

\subsection{Fault Tolerance Scalability}

\begin{figure}[t]
\Large
\centering
\begin{minipage}{.23\textwidth}\centering
\begin{tikzpicture}[scale=0.49]
\begin{axis}[
    xlabel={Throughput [ktrans/sec]},
    ylabel={Latency [ms]},
    xmin=0, xmax=21,
    ymin=0, ymax=400,
    xtick={0,5,10,15,20},
    ytick={100,200,300},
    legend style={at={(axis cs:0,400)},anchor=north west}, 
    ymajorgrids=true,
    grid style=dashed,
]

\addplot[
    color=green,
    mark=square,
    mark size=4pt,
    line width=0.5mm,
    ]
    coordinates {
    (0.078,8)(0.521,18)(3.271,32)(6.081,43)(9.203,77)(12.772,162)(16.382,321)(16.521,370)};

\addplot[
    color=blue,
    mark=+,
    mark size=4pt,
    line width=0.5mm,
    ]
    coordinates {
  (0.713,6)(6.602,30)(10.413,92)(13.413,171)(16.812,323)(17.052,372)};

\addplot[
    color=black,
    mark=star,
    mark size=4pt,
    line width=0.5mm,
    ]
    coordinates {
  (0.671,6)(6.871,25)(10.381,73)(13.893,140)(16.932,263)(17.441,291)};

\addplot[
    color=violet,
    mark=o,
    mark size=4pt,
    line width=0.5mm,
    ]
    coordinates {
    (0.872,5)(7.513,16)(12.110,64)(15.409,105)(19.105,246)(19.242,282)};

\addlegendentry{AHL}
\addlegendentry{SharPer}
\addlegendentry{coordinator}
\addlegendentry{optimistic}

\end{axis}
\end{tikzpicture}

{\footnotesize (a) $|p|=7$}
\end{minipage}\hspace{2em}
\begin{minipage}{.23\textwidth} \centering
\begin{tikzpicture}[scale=0.49]
\begin{axis}[
    xlabel={Throughput [ktrans/sec]},
    ylabel={Latency [ms]},
    xmin=0, xmax=20,
    ymin=0, ymax=400,
    xtick={0,5,10,15},
    ytick={100,200,300},
    legend style={at={(axis cs:0,400)},anchor=north west}, 
    ymajorgrids=true,
    grid style=dashed,
]

\addplot[
    color=green,
    mark=square,
    mark size=4pt,
    line width=0.5mm,
    ]
    coordinates {
    (0.069,9)(0.488,20)(3.071,33)(5.773,45)(8.661,93)(12.013,192)(14.522,290)(14.601,331)};

\addplot[
    color=blue,
    mark=+,
    mark size=4pt,
    line width=0.5mm,
    ]
    coordinates {
  (0.701,9)(6.011,60)(10.014,153)(11.892,232)(12.893,280)(13.054,341)};

\addplot[
    color=black,
    mark=star,
    mark size=4pt,
    line width=0.5mm,
    ]
    coordinates {
  (0.633,6)(6.417,25)(9.802,78)(13.190,161)(16.001,272)(16.165,320)};

\addplot[
    color=violet,
    mark=o,
    mark size=4pt,
    line width=0.5mm,
    ]
    coordinates {
    (0.843,4)(7.119,16)(11.503,69)(14.772,110)(18.013,247)(18.114,301)};

\addlegendentry{AHL}
\addlegendentry{SharPer}
\addlegendentry{coordinator}
\addlegendentry{optimistic}
 
\end{axis}
\end{tikzpicture}

{\footnotesize (b) $|p|=13$}
\end{minipage}
\caption{Increasing the Number of Byzantine Nodes}
  \label{fig:nodes-b}
\end{figure}
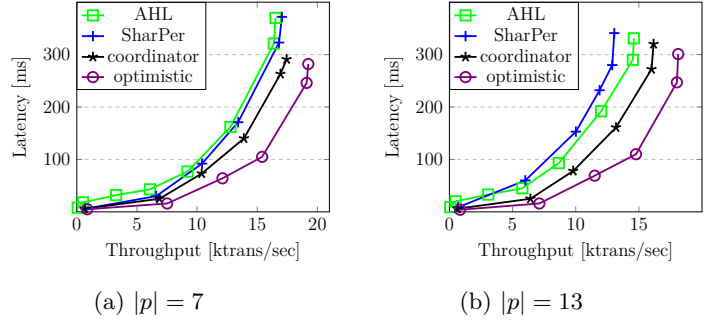

Finally, we evaluate the impact of increasing the number of nodes within each domain on the performance of protocols.
Figures ~\ref{fig:nodes-c} and \ref{fig:nodes-b} depict the results.
We consider two scenarios with $f=2$ and $f=4$, i.e.,
each crash-only domain includes $5$ and $9$ nodes, and
each Byzantine domain includes $7$ and $13$ nodes respectively.
All nodes are placed within an AWS region and the workload includes
$90\%$-internal $10\%$-cross-domain transactions.
When domains become larger, achieving consensus requires more nodes, hence,
the performance of all protocols is (marginally) reduced, e.g.,
the throughput of the coordinator-based protocol is reduced by $6\%$ and $11\%$ (with the same latency)
when the size increases from $3$ to $5$ and $9$.

\subsection{Evaluation Summary}
Overall, the evaluation results can be summarized as follow.
First, the coordinator-based protocol outperforms SharPer and AHL, demonstrating a scalable solution that can be practically
deployed over wide-area networks and used for all types of workloads.
Second, in low contention workloads, the optimistic protocol processes transactions efficiently because it does not require communication across domains.
However, in high contention workloads, the protocol performance is significantly reduced due to inconsistency between the ledgers of different domains, which leads to aborting all their data-dependent transactions.
Third, while SharPer outperforms AHL in nearby domains,
AHL demonstrates better performance in far apart domains due to its coordinator-based consensus protocol.
Finally, \sys supports mobility over wide-area networks efficiently.
\section{Related Work}\label{sec:related}

Despite several years of intensive research, existing blockchain solutions
do not adequately address the performance and scalability requirement of edge computing networks,
which is characterized by cross-domain transactions and possibly mobile nodes communicating over wide-area networks.
\ifextend
In general, the ordering and execution of transactions are the two main phases of processing transactions.
While early permissioned blockchains, e.g., Tendermint \cite{kwon2014tendermint} and Quorum \cite{morgan2016quorum},
use the sequential order-execute paradigm, 
Hyperledger Fabric \cite{androulaki2018hyperledger}
improves performance by leveraging parallel execution of transactions.
Several recent permissioned blockchains, e.g.,
blockchain relational database \cite{nathan2019blockchain},
ParBlockchain \cite{amiri2019parblockchain},
Fast Fabric \cite{gorenflo2019fastfabric}, 
XOX Fabric  \cite{gorenflo2020xox},
Fabric++ \cite{sharma2019blurring}, and
FabricSharp \cite{ruan2020transactional}
also, execute transactions in parallel and try to address the shortcomings of Hyperledger Fabric,
e.g., dealing with contentious workloads.
\fi

Processing globally distributed transactions across multiple clusters (e.g., data centers)
have been discussed in several studies~\cite{baker2011megastore}\cite{bronson2013tao}\cite{corbett2013spanner}\cite{decandia2007dynamo}\cite{glendenning2011scalable}\cite{kallman2008h}\cite{mahmoud2013low}\cite{patterson2012serializability}\cite{prasaad2020handling}\cite{taft2014store}\cite{thomson2012calvin}\cite{zamanian2020chiller}.
These systems typically shard data and replicate each data shard on multiple clusters.
A coordinator-based approach, e.g., two-phase commit and two-phase locking, is then used for cross-cluster communication
while a crash fault-tolerant protocol, e.g., Paxos, is used to guarantee fault tolerance within each cluster. 
The coordinator-based protocol of \sys is different from all these systems in three ways. 
First, in \sys, nodes might follow Byzantine failure model.
Second, \sys sacrifices availability for performance by replicating data only on one (nearby) domain, and
third, \sys leverages the hierarchical structure of edge computing networks to rely on the lowest common ancestor of all involved domains
to play the coordinator role.

Processing distributed transactions across multiple clusters in the presence of Byzantine nodes has also been addressed in several studies, e.g., permissioned blockchains.
Partially replicated systems replicate each data shard on a single cluster and use either coordinator-based protocols, e.g., AHL \cite{dang2018towards},
or flattened protocols, e.g., SharPer \cite{amiri2019sharding}\cite{amiri2021sharper}, to process cross-shard transactions.
However, sharding approaches maintain data shards mainly on cloud servers with possibly large network distances from edge devices.
Moreover, the far network distance either between the involved shards (in the flattened approach) or
between the coordinator and involved shards (in the coordinator-based approach)
results in high latency.

In geo-distributed fully replicated systems, e.g.,
Steward \cite{amir2008steward}, Blockplane \cite{nawab2019blockplane}, and GeoBFT \cite{gupta2020resilientdb},
the data is replicated on every cluster.
GeoBFT proceeds in rounds where at every round, each cluster establishes consensus on a transaction and
multicasts the locally-replicated transaction to other clusters.
All clusters then, execute all transactions of that round in a predetermined order.
Blockplane and Steward, on the other hand,
present a hierarchical two-level approach where different clusters locally establish BFT consensus on disjoint transactions, and
at the top level, a CFT consensus protocol is used to process all transactions globally.
In contrast to geo-distributed systems, \sys assumes the geographical locality of data access
(a reasonable assumption in edge commuting networks) and replicates data only on a nearby domain.
While this design choice brings down the availability guarantee of \sys, it leads to higher performance.
Furthermore, \sys leverages the hierarchical structure of edge computing networks to process cross-domain transactions more efficiently.

\ifextend
The sharding technique has been used in both permissionless,
e.g., Elastico \cite{luu2016secure}, OmniLedger \cite{kokoris2018omniledger} and Ethereum 2 \cite{Shardchains},
and permissioned blockchains,
e.g., AHL \cite{dang2018towards}, Chainspace \cite{al2017chainspace}, SharPer \cite{amiri2021sharper} and
Blockplane \cite{nawab2019blockplane}.
However, sharding has several shortcomings that make it inappropriate to be used in wide area networks.
First, sharding approaches maintain data shards mainly on centralized cloud servers which can have large network distances from edge devices. Traditional sharding assumes a flat hierarchy and does not leverage the inherent hierarchical structure of the Internet to optimize for low latency communication.
Second, sharding either use a flattened approach, e.g., SharPer \cite{amiri2021sharper},
or a coordinator-based approach, e.g., AHL \cite{dang2018towards} to process cross-shard transactions where
in both approaches, as explained earlier,
the far network distance either between the involved shards (in the flattened approach) or
between the coordinator and involved shards (in the coordinator-based approach)
results in high latency.
Third, while many edge devices in wide area networks are mobile,
mobility has not been addressed in sharding solutions.

GeoBFT \cite{gupta2020resilientdb}, as another scalability solution,
maintains the entire ledger on every node
and orders each transaction within only a single cluster
to reduce the latency of cross-cluster transactions.
After ordering every transaction, however,
all clusters need to communicate with each other 
to enable other clusters to update their ledgers and execute transactions,
resulting in high latency.
\fi

The blockchain model presented in \cite{sahoo2019hierarchical}
focuses only on the data abstraction across different levels of the hierarchy and
does not address cross-domain transactions, consensus, and mobility of nodes.
Plasma \cite{poon2017plasma} also uses
hierarchical chains to improve transaction throughput of the Ethereum blockchain,
however, processing cross-domain transactions and mobility of nodes
have not been addressed in Plasma. 
\ifextend
Our work is also related to blockchain systems with DAG-structured ledgers, e.g.,
Iota \cite{popov2016tangle}, Vegvisir \cite{karlsson2018vegvisir},
Phantom \cite{sompolinsky2018phantom}, Spectre \cite{sompolinsky2016spectre}, and Caper \cite{amiri2019caper}.
While in such systems, the DAG structure is the result of
the simultaneous appending of transactions to the ledger,
in \sys, the DAG structure is used to support cross-domain transactions in internal levels.
In particular, in Caper, the DAG structure is the result of
the simultaneous appending of local and global transactions to the ledger whereas,
in \sys, the DAG structure is used to support cross-domain transactions in internal levels.
Moreover, Caper neither follows the hierarchical structure of wide area networks
nor supports the mobility of nodes.
\fi

Blockchain brings the capability of managing edge computing network data through
its secure distributed ledger and
provide immutability, decentralization, and transparency, all of which promise
to tackle privacy and security challenges of current edge computing networks
\cite{alaslani2019blockchain}\cite{dorri2016blockchain}\cite{guo2019blockchain}\cite{nguyen2020blockchain}\cite{singh2019wedgedb}\cite{yuan2021coopedge}.
Nonetheless, these studies do not address the challenges of maintaining hierarchical ledgers,
processing cross-domain transactions,
and consensus with mobile devices.
\ifextend
Finally, in comparison to mobile databases \cite{imielinski1992querying,barbara1999mobile},
while \sys has a similar architecture,
the trust model (e.g., Byzantine nodes), the level of mobility (only edge vs. every node), and how \sys
establishes consensus and tracks mobility are different.
\fi
\section{Conclusion}\label{sec:conc}

In this paper, we present \sys, a permissioned blockchain system that
leverages the hierarchical structure of edge computing networks to achieve four main purposes.
First, \sys processes cross-domain transactions using a coordinator-based approach
by relying on the lowest common ancestor of the involved domains.
Second, in \sys domains propagate (a summarized version of) their ledgers up the hierarchy to provide data aggregation functionalities.
Third, \sys presents an optimistic cross-domain consensus protocol by relying on higher-level nodes to detect inconsistencies.
Finally, \sys addresses the mobility of edge devices by introducing a mobile consensus protocol.
We validated these technical innovations by developing a prototype of \sys,
where our evaluation results across a wide range of workloads demonstrate
the scalability of \sys in processing cross-domain transactions across edge computing networks and
transactions initiated by mobile devices where the involved nodes are far apart.
\balance

\bibliographystyle{abbrv}
\bibliography{_blockchain,_misc,_privacy,_system}

\end{document}